\documentclass[preprint,prd,aps,tighten,nofootinbib,amssymb]{revtex4}
\usepackage{color}
\input{colordvi.tex}
\usepackage{amsmath}
\usepackage[dvips]{graphicx}
\usepackage{subfigure}
\newcommand{\beq}{\begin{equation}}
\newcommand{\eeq}{\end{equation}}
\newcommand{\beqa}{\begin{eqnarray}}
\newcommand{\eeqa}{\end{eqnarray}}

\newcommand{\CO}{{\cal O}}

\newcommand{\bea}{\begin{eqnarray}}
\newcommand{\eea}{\end{eqnarray}}
\newcommand{\bear}{\begin{array}}
\newcommand {\eear}{\end{array}}
\newcommand{\bef}{\begin{figure}}
\newcommand {\eef}{\end{figure}}
\newcommand{\bec}{\begin{center}}
\newcommand {\eec}{\end{center}}

\begin{document}
\widetext
\draft

\begin{flushright}
RIKEN-MP-50\\
DESY 12-124\\
TU-915 
\end{flushright}

\title{Higgs, Moduli Problem, Baryogenesis \\
and Large Volume Compactifications
}

\author{Tetsutaro Higaki}%
\affiliation{Mathematical Physics Lab., RIKEN Nishina Center, Saitama 351-0198, Japan}
\author{Kohei Kamada}%
\affiliation{Deutsches Elektronen-Synchrotron DESY, Notkestra\ss e 85, D-22607 Hamburg, Germany}
\author{Fuminobu Takahashi}
\affiliation{Department of Physics, Tohoku University, Sendai 980-8578, Japan}

\date{\today}

\begin{abstract}
We consider the cosmological moduli problem in the context of high-scale supersymmetry breaking suggested by the recent
discovery of the standard-model like Higgs boson. In order to solve the notorious moduli-induced gravitino problem,
we focus on the LARGE volume scenario, in which the modulus decay into gravitinos can be kinematically forbidden.
We then consider the Affleck-Dine mechanism with or without an enhanced coupling
with the inflaton, taking account of possible Q-ball formation.
We show that the baryon asymmetry of the present Universe can be
generated  by the Affleck-Dine mechanism in LARGE volume scenario, solving the moduli and gravitino problems.
We also find that the overall volume modulus decays into a pair of the axionic
superpartners, which contribute to the extra relativistic degrees of freedom. 

\end{abstract}

\maketitle
\section{Introduction}
\label{sec:1}

Recently the ATLAS and CMS collaborations have  discovered
 a standard model (SM)-like Higgs particle with mass of about
$125$\,GeV~\cite{JULY4thtalk} (see also \cite{ATLAS:2012ae,Chatrchyan:2012tw}).  The relatively light Higgs boson mass strongly
suggests the presence of new physics at a scale below the Planck scale~\cite{EliasMiro:2011aa}.
From both phenomenological and theoretical points of view, supersymmetry (SUSY) is arguably the  
most plausible candidate for  the new physics beyond the SM. 

In the minimal supersymmetric extension of the SM,  the 125 GeV Higgs mass can be explained without invoking 
large stop mixing if the typical sparticle mass is at $O(10)$\,TeV or heavier. In particular,
the SUSY should appear at a scale below PeV for $\tan \beta \gtrsim 2$,  where $\tan \beta$ represents
the ratio of the up-type and down-type Higgs boson vacuum expectation values (VEVs). 
It is therefore of utmost importance to study the cosmological and phenomenological implications of
such high-scale SUSY suggested by the Higgs boson mass.

The string theory is a plausible candidate for an underlying high-energy theory, and in particular,
it seems to possess some features the quantum theory of gravity should possess.
However, it suffers from a serious cosmological moduli problem~\cite{Coughlan:1983ci}.
Massless moduli fields parametrize the continuous ground state degeneracies and they 
generally appear in the compactifications of extra dimensions.  In order to construct 
phenomenologically viable models, those moduli fields need to be stabilized. However, some of them remain 
relatively light, acquiring masses  induced only by the SUSY breaking. Because of its light mass,
those moduli are copiously produced after inflation as coherent oscillations. 
If the moduli mass is of order the weak scale or lighter, they  typically decay after the big bang nucleosynthesis (BBN),
thus altering the standard cosmology in contradiction with observations.

In the case of high-scale SUSY as suggested from the SM-like Higgs boson mass, 
the moduli fields with mass heavier than $100$\,TeV or so
decay before the BBN,  and the cosmological moduli problem is greatly relaxed. 
However the moduli generically decay into gravitinos with a sizable branching fraction if kinematically allowed,
and  those gravitinos produce  lightest SUSY particles (LSPs), whose abundance easily 
exceeds the observed dark matter density.  This is known as the moduli-induced
gravitino problem~\cite{moduli,Dine:2006ii,Endo:2006tf}. Furthermore, a huge amount of entropy is produced 
by the modulus decay, and any pre-existing baryon asymmetry would be diluted by a significant factor. 
Therefore it is difficult to generate the right amount of the
baryon asymmetry via   the standard leptogenesis~\cite{Fukugita:1986hr},  and we need more efficient baryogenesis
such as the Affleck-Dine (AD) mechanism~\cite{Affleck:1984fy}.  Thus, solving the issues of the moduli and gravitino problems and
the origin of the baryon asymmetry  is the key to understand the evolution of the Universe and high-energy theory.

Our strategy is twofold. First we revisit the AD mechanism in high-scale SUSY breaking, 
taking account of possible Q-ball formation~\cite{Coleman,Kusenko97}. 
We will show that the Q balls decay sufficiently fast and so they are cosmologically harmless,
both because of the high-scale SUSY breaking, and because of the mild hierarchy between the gaugino mass 
and the scalar mass. 
Secondly we consider the moduli problem in a realistic moduli stabilization. In order to solve the serious moduli-induced 
gravitino problem, we consider the LARGE volume scenario (LVS)~\cite{Balasubramanian:2005zx}, 
in which the modulus decay into gravitinos can be 
kinematically forbidden. Interestingly, the cut-off scale tends to be smaller than the Planck scale in LVS, 
and if the coupling between the inflaton and the AD field is enhanced, 
the resultant baryon asymmetry increases significantly. We will show that the right amount of the baryon asymmetry can be
naturally generated by the AD mechanism in LVS, solving the moduli and gravitino problems.

The rest of the paper is organized as follows. In Sec.~\ref{sec:2} we consider the AD mechanism
in high-scale SUSY breaking to study if it can generate a sufficient amount of the baryon asymmetry in the
presence of a huge entropy production by the modulus decay. We give two concrete realizations of
the moduli stabilization and discuss the cosmological issues in Sec.~\ref{sec:3} and Sec.~\ref{sec:4}.
The last section is devoted for discussion and conclusions.

\section{Affleck-Dine baryogenesis}
\label{sec:2}

We first review the AD mechanism \cite{Affleck:1984fy,Dine:1995uk}, in which the 
baryon asymmetry is generated thorough the dynamics of scalar fields with baryon and/or lepton charge.
To be concrete we assume a mild hierarchy between gaugino and scalar masses, such that the scalar
mass is several orders of magnitude heavier than the typical gaugino mass. 
Such mass spectrum is realized in simple anomaly mediation with a generic K\"ahler potential as well
as in the modulus mediation, as we shall see in the next section.

We consider the effect of a large negative Hubble-induced mass 
on the AD mechanism and its cosmological consequences. Such an enhanced coupling of the AD field
with the inflaton is expected in the context of LVS. 
We will see that the  baryon asymmetry of the present Universe can be explained, even in the presence
of a huge entropy production by the modulus decay. In particular, the Q balls decay before BBN, and so,
they do not play any important role in our scenario.

\subsection{Affleck-Dine baryogenesis in high-scale SUSY breaking}
In supersymmetric theories,  flat directions are ubiquitous~\cite{hep-ph/9510370}.
The scalar potential along  flat directions vanishes in the exact SUSY limit at renormalizable level. 
A flat direction can be parameterized by a  gauge invariant monomial  such as 
$udd$ or $L H_u$, and its dynamics can be described in terms of a complex scalar field $\phi$, 
which we call the AD field. 
If the AD field $\phi$ carries baryon and/or lepton charge, its dynamics 
can generate the baryon asymmetry in the Universe. Hereafter we assume that 
the AD field $\phi$ carries a non-vanishing baryon charge, $\beta$.

The flat direction is lifted by both non-renormalizable interactions and SUSY breaking. 
We consider a non-renormalizable superpotential of the following form, 
\begin{equation}
W=\frac{y}{M_*^{n-3}}\phi^n, 
\end{equation}
where $y$ is a coupling constant, $M_*$ is the cut-off scale, and $n$ is an integer greater than $3$.
We set $y$ to be real and positive without loss of generality. The value of $n$ depends on  flat directions
as well as on the existence of a possible discrete symmetry under which $\phi$ is charged. 
Taking account of soft SUSY-breaking effects,  the scalar potential in a flat space time is expressed as, 
\begin{equation}
V(\phi)= m_0^2 |\phi|^2 + \left(\frac{A_n y}{M_*^{n-3}}\phi^n+ {\rm h.c.} \right) + \frac{n^2y^2}{M_*^{2n-6}}|\phi|^{2n-2},
\end{equation}
where $m_0$ and $A_n$ are the soft scalar mass of $\phi$ and the coefficient of the $A$-term, respectively.
The magnitude of $m_0$ and $A_n$ depend on the  SUSY-breaking mediation mechanism. 
The $A$-term violates the baryon number explicitly, which is the source of baryon asymmetry.
Hereafter we assume that $m_0$ is of order $100$\,TeV or heavier,  in order to explain the SM-like Higgs boson
with mass about $125$\,GeV. As a reference value we will set $m_0 = 10^{3}$\,TeV in the following analysis.

Now we consider the dynamics of the AD field in the inflationary Universe. 
In the case where inflation is driven by the $F$-term of a canonically normalized inflaton, $I$, 
the AD field generally acquires a mass squared whose amplitude is of order the Hubble parameter squared, through
 the Planck-suppressed interaction in supergravity. 
We here assume that the sign of the Hubble-induced mass is negative\footnote{
If the sign of the Hubble induced mass is positive, the AD field settles down to the origin 
during inflation and it does not play any important role in cosmology. }, 
\begin{equation}
V_H=-c^2 H^2 |\phi|^2, 
\end{equation}
where  $H$ is the Hubble parameter and 
$c$ is a positive numerical coefficient.  For the moment, 
we consider the case of $c=O(1)$. Such a negative mass term is generated if there is a quartic 
coupling between inflaton and the AD field in the K${\rm {\ddot a}}$hler potential,
\begin{equation}
K_{\rm NM}=\frac{a}{M_{\rm pl}}|\phi|^2 |I|^2, \label{plkal}
\end{equation}
where $a$ is a positive numerical coefficient of order unity and $M_{\rm pl}$ is 
the reduced Planck mass.

Suppose that the Hubble-induced mass $cH$ during inflation is larger than the soft scalar mass $m_0$.
Then, the origin of $\phi$ is  destabilized  and it settles down at the potential minimum determined by 
the balance between the negative Hubble-induced mass 
and the $F$-term from the non-renormalizable superpotential, 
acquiring a large expectation value:
\begin{equation}
|\phi_{\rm inf}| \simeq \left(\frac{c H_{\rm inf} M_*^{n-3}}{n\sqrt{n-1}y}\right)^{1/(n-2)}, 
\end{equation} 
where $H_{\rm inf}$ is the Hubble parameter during inflation.  Such initial condition is
one of the requisites for the AD mechanism, namely, 
\begin{equation}
H_{\rm inf} > \frac{m_0}{c}, 
\label{conditionforAD}
\end{equation}
is required.

Let us study the dynamics of the AD field $\phi$ after inflation and see how the baryon asymmetry is generated. 
In the single-field inflation, $\phi$ still receives a negative Hubble-induced mass after inflation,  as long as the energy density of the Universe 
is dominated by the inflaton matter. In the case of multi-field inflation, we assume that the Hubble-induced mass term is negative and its
magnitude does not change significantly during and after inflation. After inflation, $\phi$ then follows the time-dependent minimum $(cHM_*^{n-3}/n\sqrt{n-1}y)^{1/(n-2)}$ until 
$\phi$ starts oscillations around the origin when $H\simeq m_0/c$.\footnote{Here we assume that the thermal correction to the scalar potential 
does not affect the dynamics of $\phi$. This is considered to be the case if the field value of $\phi$ is sufficiently large.}. 
At the onset of oscillations,  $\phi$  is kicked into the phase direction by the $A$-term, and the 
baryon number is generated. The resultant baryon number density,
\begin{equation}
n_B=i \beta ( \phi {\dot \phi}^* - {\dot \phi} \phi^*), 
\end{equation}
can be evaluated by solving the equation of evolution, 
\begin{equation}
{\dot n}_B + 3H n_B=2 \beta {\rm Im} \left(\frac{\partial V}{\partial \phi} \phi \right)=\frac{2n\beta A_n y}{M_*^{n-3}} {\rm Im} (\phi^n). \label{eqforbar}
\end{equation}
Assuming that the scale factor evolves as $a\propto t^{2/3}$ during inflaton matter domination, 
we obtain the resultant baryon number density at $t > t_{\rm osc}$:
\begin{align}
n_B (t)&= \frac{2\beta}{a^3(t)} \int^{t} dt^\prime a^3(t^\prime) {\rm Im} \left(\frac{\partial V}{\partial \phi} \phi \right) \notag \\
&\simeq \left(\frac{a(t_{\rm osc})}{a(t)}\right)^3 \frac{2(n-2)}{3n^{2/(n-2)}(n-1)^{n/(2n-4)}(n-3)} \frac{\beta c^{n/(n-2)}}{y^{2/(n-2)}} A_n M_*^{2(n-3)/(n-2)} H_{\rm osc}^{2/(n-2)} \delta_{\rm eff} \notag \\
&\simeq   \left(\frac{a(t_{\rm osc})}{a(t)}\right)^3 \frac{2(n-2)}{3\sqrt{n-1}(n-3)} \beta c A_n \phi_{\rm osc}^2 \delta_{\rm eff} , 
\end{align}
where the subindex `osc' means that the variable is evaluated at the onset of the oscillations.
Here $\delta_{\rm eff} \leq 1$ is the CP phase factor, and it is typically of $\CO(0.1)$,  without fine-tunings of
 the initial phase of the AD field.
The baryon asymmetry in a comoving volume $a^3 n_B$ is conserved soon after the onset of the AD field oscillation, 
because the oscillation amplitude of $\phi$  decreases rapidly due to the Hubble friction.

So far we have assumed that the oscillation in the phase direction does not begin until the onset of the oscillation in the radial 
direction. This assumption is valid as long as the $A$-term is small enough, 
\begin{equation}
|A_n| < \frac{\sqrt{n-1}}{n^2 c^2} m_0 \equiv A_n^{\rm cri}. 
\label{condAterm}
\end{equation}
If the AD field starts to oscillate in the phase direction earlier, 
the estimate of the resultant baryon asymmetry becomes a little more complicated. 
Dividing the AD field into the radial and phase component, $\phi=|\phi|e^{i \theta}$, 
the equation of motion for the phase component reads, 
\begin{equation}
{\ddot \theta}+\left(3H+2\frac{|{\dot \phi}|}{|\phi|}\right) {\dot \theta}+\frac{A_n y}{M_*^{n-3}}|\phi|^{n-2}\sin (n \theta)=0. 
\end{equation}
Substituting the expression $|\phi|\sim (c H M_*^{n-3}/\sqrt{n-1}y)^{1/(n-2)}$,
this equation can be approximated as
\begin{equation}
{\ddot \theta}+\frac{2(n-3)}{(n-2)t}{\dot \theta}+\frac{2n c A_n}{3\sqrt{n-1} t} \theta=0. \label{phaseeom}
\end{equation}
Here we have assumed that $\theta$ is not so large that one can approximate $\sin ( n \theta) \sim n \theta$ 
and used $H=2/3t$. 
Solving Eq. \eqref{phaseeom} for $n=6$, 
we obtain an approximate solution  after the onset of the oscillation along the 
phase direction,
\begin{equation}
\theta(t)\simeq \sqrt{\frac{t_{\rm ph}}{t}}\cos\left(2\sqrt{\frac{t}{t_{\rm ph}} } \right)\theta_0,
\end{equation}
where $t_{\rm ph} = (2ncA_n/3\sqrt{n-1})^{-1}$ is the time at the onset of the oscillation in the phase direction and 
$\theta_0$ is the AD field phase at $t=t_{\rm ph}$. 
Therefore, the time derivative of $\theta$ at the onset of its oscillation along the radial direction is estimated as
\begin{equation}
{\dot \theta} (t_{\rm osc})\sim \frac{m_0}{c} \delta_{\rm eff},  
\end{equation}
where we include the overall CP phase factor $\delta_{\rm eff}=\theta_0\, \delta_{\rm osc}$. Here $\delta_{\rm osc}$
represents how efficient the phase velocity turns into the baryon asymmetry at $t = t_{\rm osc}$.
Then, we can approximate the baryon number density at the onset of the AD field oscillation in the radial direction, 
\begin{equation}
n_B(t_{\rm osc})\simeq \beta \phi_{\rm osc}^2 {\dot \theta} \sim \frac{m_0}{c} \phi_{\rm osc}^2\delta_{\rm eff}.  
\end{equation}
Similar to the former case, the baryon asymmetry in the comoving volume is fixed soon after the onset of the 
AD field oscillation in the radial direction. 
As a result, we arrive at an approximate estimation, 
\begin{equation}
n_B(t)\simeq   \left(\frac{a(t_{\rm osc})}{a(t)}\right)^3 \beta c A_n \phi_{\rm osc}^2 \delta_{\rm eff}  \times  
f
\end{equation}
with
\beq
f\equiv\left\{
\begin{array}{ll}
1 & \text{for } \quad A_n<A_n^{\rm cri}, \\
\dfrac{A_n^{\rm cri}}{A_n}& \text{for } \quad A_n>A_n^{\rm cri}, 
\end{array}\right. 
\end{equation}
where we have omitted an $\CO(1)$ numerical factor. 
Note that there is an upper bound on $c$, 
\begin{equation}
c<\frac{\sqrt{n-1}H_{\rm inf}}{n A_n}, 
\end{equation}
in order for the AD field not to settle down to the potential minimum in the phase direction during inflation. 
Moreover, for the case $c\sim \CO(1)$, $A_n$ should not be much larger than $A_n^{\rm cri}$ 
since there arises a color-breaking potential minimum other than the origin in the scalar potential.

The present baryon-to-entropy ratio is then evaluated as 
\begin{align}
\frac{n_B}{s}(t_0)&\simeq \Delta^{-1} \frac{n_B}{s}(t_R) \\
& \simeq \frac{n(n-2)}{6 \sqrt{n-1}(n-3)}\beta c^3 f \delta_{\rm eff} \Delta^{-1} \frac{A_n T_R}{m_0^2} \left(\frac{\phi_{\rm osc}}{M_{\rm pl}}\right)^2  \notag \\
& \sim  10^{-10} c^3 f \delta_{\rm eff} \left(\frac{\Delta^{-1}}{10^{-3}}\right)\left(\frac{A_n}{10^3 {\rm GeV}}\right) \left(\frac{T_R}{10^7 {\rm GeV} } \right)\left(\frac{m_0}{10^6{\rm GeV}}\right)^{-2} \left(\frac{\phi_{\rm osc}}{10^{16}{\rm GeV}}\right)^2,
\end{align}
where 
$T_R$   is the reheating temperature. 
Here we have inserted the entropy dilution factor $\Delta$ due to the modulus  decay.  
Note that if there is a moduli dominated era,  the baryon asymmetry is diluted by a factor $\Delta^{-1}\simeq T_X/T_{\rm dom}$ 
where the subindices $X$ and `dom' represent that the variables are evaluated at the 
modulus decay and at the onset of the moduli domination, respectively.

Now we turn to the cosmological effect of Q balls \cite{Coleman,Kusenko97}, which are potential obstacles in this scenario. 
A Q ball is a non-topological soliton, and the Q-ball solution exists when the scalar potential is flatter than the quadratic potential, 
and its stability is guaranteed  by the conserved (baryon and/or lepton) charge.
In the AD mechanism, this condition on the scalar potential is met  if the radiative correction 
to the soft mass is negative~\cite{Kusenko97, Enqvist98}, i.e., 
\begin{equation}
V(\phi)\ni m_0^2 |\phi|^2 \left(1+K \log \left(\frac{|\phi|^2}{\Lambda^2}\right)\right), 
\end{equation}
with $K<0$.
Here $\Lambda$ is the SUSY-breaking scale where $m_0$ is evaluated. 
$K$ is a numerical coefficient of the one-loop
radiative corrections and it becomes negative 
when the gaugino loop dominates. 
For example, in case of the $udd$ flat direction,  $K$ is evaluated as  \cite{Enqvist98}
\begin{equation}
K \simeq -\frac{4\alpha_3}{3\pi}  \frac{M_3^2}{m_0^2} 
\simeq  10^{-6} \times \left(\frac{M_3}{10^4 {\rm GeV}}\right)^2 \left(\frac{m_0}{10^6{\rm GeV}}\right)^{-2}, 
\end{equation}
where we have considered only gluino contributions. 
Note that we consider a case where there is a hierarchy between the scalar mass $m_0$ and the gaugino mass $M_a$. 
For example, in the anomaly mediation with a general K${\rm {\ddot a}}$hler potential, 
the hierarchy  is given by $M_a/m_0\sim M_a/m_{3/2} \sim g_a^2/16\pi^2= 10^{-3} - 10^{-2}$. 
As a result, $|K|$ is suppressed compared to the case of $m_0 \simeq M_a$\footnote{
It is possible that top (and bottom) loop contributions make $K$ positive, 
depending on the value of tan $\beta$. In this case Q balls are not formed.}.

If $K$ is negative, the AD field condensate experiences spatial instabilities and   Q balls  are formed at about $H=H_* \sim 0.1 m_0 |K|$.
The Q balls have the  following properties \cite{Enqvist98, kasuya00}, 
\begin{equation}
R \simeq |K|^{-1/2} m_0^{-1}, \quad \omega \simeq m_0, \quad \phi_Q \simeq \frac{H_*}{H_{\rm osc}} \phi_{\rm osc} \simeq 0.1 c|K| \phi_{\rm osc},  \label{qprop}
\end{equation}
where $R$ is the radius of a Q ball, $\omega$ is the angular momentum of the AD field inside a Q ball, and $\phi_Q$ is the field value 
at the center of a Q ball. 
The charge and the energy stored in a Q ball are estimated as 
\begin{equation}
Q \simeq R^3 \omega \phi_Q^2\simeq 10^{15} c^2 \left(\frac{|K|}{10^{-6}}\right)^{1/2} \left(\frac{m_0}{10^6{\rm GeV}}\right)^{-2} \left(\frac{\phi_{\rm osc}}{10^{16} {\rm GeV}}\right)^2, \quad E_Q \simeq m_0 Q. 
\end{equation}

If the Q balls are stable, they may overclose the Universe. Alternatively, if they are unstable and decay into lighter degrees of freedom
during BBN, it may change the light element abundances spoiling the success of the BBN.
In the present case,  however, the energy per unit charge, $E/Q$, is comparable to the soft scalar mass, $m_0$, and hence they can 
decay into quarks and lighter SUSY particles.
The decay proceeds from the Q-ball surface and its rate is evaluated as \cite{Cohen:1986ct}
\begin{align}
\Gamma_Q&=\frac{1}{Q} \frac{dQ}{dt} \simeq \frac{\omega^3 R^2}{48 \pi Q} \\
&\simeq 10^{-5}{\rm GeV} c^{-2}\left(\frac{|K|}{10^{-6}}\right)^{-3/2} \left(\frac{m_0}{10^6{\rm GeV}}\right)^{3} \left(\frac{\phi_{\rm osc}}{10^{16}{\rm GeV}}\right)^{-2}. \label{qdecay}
\end{align}
Thus, the decay temperature of the Q balls is 
\begin{equation}
T_{\rm dec}\simeq \sqrt{\Gamma_Q M_{\rm pl}}\simeq 5 \times 10^6{\rm GeV} c^{-1}\left(\frac{|K|}{10^{-6}}\right)^{-3/4} \left(\frac{m_0}{10^6{\rm GeV}}\right)^{3/2} \left(\frac{\phi_{\rm osc}}{10^{16}{\rm GeV}}\right)^{-1} . 
\end{equation}
Therefore, the Q balls decay much before BBN even if 
$\phi_{\rm osc}$ is as large as the Planck scale, both because of
the heavy scalar mass and because of the small value of $|K|$. 
The Q balls are cosmologically harmless in the high-scale SUSY breaking.

\subsection{Effect of large negative Hubble induced mass}

So far, we have assumed that the negative mass of the AD field during and after inflation 
is of  order the Hubble parameter, i.e., $c \simeq \CO(1)$.  
Here let us consider the case where  the coupling of the inflaton and the AD field 
 in the K${\rm {\ddot a}}$hler potential  is enhanced,
 \begin{equation}
K_{\rm NM2}=\frac{1}{{\tilde M}^2}|\phi|^2|I|^2, \quad {\tilde M} \ll M_{\rm pl}.
\label{Keff}
\end{equation}
Then
the large negative Hubble-induced mass is generated,
\begin{equation}
V_{\rm H} = - c^2 H^2 |\phi|^2, \quad c \simeq \frac{M_{\rm pl}}{{\tilde M}} ,
\end{equation}
where $\tilde M$ denotes the effective cut-off scale for the enhanced coupling. 
Such an enhancement is indeed realized in the string inspired models as we shall see later.

The large negative Hubble induced mass delays the onset of the AD field oscillation.
Then, the Hubble friction  is suppressed when the AD field starts to oscillate.
One may worry  that,  since the AD field oscillates many times in one Hubble time,
the AD field dynamics after the onset of its oscillations
may wash out the baryon asymmetry generated by the first kick and the resultant baryon 
asymmetry may be suppressed. 
However, because the time scale of the first kick, $H_{\rm osc} ^{-1}\simeq c m_0^{-1}$, is much longer than the typical time scale
of oscillations,  $m_0^{-1}$, the effect of  the wash out tends to be negligibly small.
As a result, the baryon asymmetry is determined by the first kick and 
becomes fixed soon after the commencement of the oscillations. 
The baryon asymmetry is therefore given by  
\begin{align}
\frac{n_B}{s}(t_0) &\sim 10^{-10} f \delta_{\rm eff} \left(\frac{c}{10}\right)^3 \left(\frac{\Delta^{-1}}{10^{-3}}\right)\left(\frac{A_n}{10^3 {\rm GeV}}\right) \left(\frac{T_R}{10^6 {\rm GeV} } \right)\left(\frac{m_0}{10^6{\rm GeV}}\right)^{-2} \left(\frac{\phi_{\rm osc}}{10^{15}{\rm GeV}}\right)^2. 
\label{YB}
\end{align}
Note that the resultant baryon-to-entropy-ratio is enhanced by a factor of $c^3$ or $c$. (Note that $f$ depends on $c$
if the A-term is large.)

The later onset of the AD field oscillation also affects the Q-ball properties. 
This results in the larger field value of the AD field at the center of Q balls and the larger charge stored in a 
Q ball.
Although the large charge of Q balls suppresses the decay rate by a factor of $c^{-2}$ (Eq.~\eqref{qdecay}), 
the Q balls can decay before BBN for a wide range of $c$. However, in order to avoid the overproduction of the LSP produced
by the Q-ball decay, 
$c$ should not be too large. Typically it should satisfy  $c \lesssim 10^3$. The precise value of the upper bound
depends on the thermal history of the Universe. If the modulus field decays later, the LSP produced
by the Q-ball decay is diluted. However, as shown in Ref.~\cite{moduli}, the moduli tends to have a sizable (not chiral suppressed)
branching fraction of the decay into gauginos. So in this case the LSPs may be still overproduced.
That said, as we shall see later, the modulus field  decays sufficiently fast in the concrete moduli stabilization because of the heavy gravitino mass.
Therefore, the upper bound on $c \lesssim 10^3$ is a more or less reasonable constraint in our scenario. 

In summary, the large negative Hubble induced mass enhances the baryon asymmetry of the Universe. 
This will help the situation of string-inspired SUSY-breaking models where 
the baryogenesis is difficult due to the late time entropy production from modulus decay. 

Furthermore, the inflation scale is generically bounded above in order not destabilize the moduli fields. 
It is therefore non-trivial if the AD mechanism works in such low-scale inflation model, because the AD
field may be stabilized at the origin during inflation.   However,  if $c \gg 1$, 
the origin can be destabilized by the large negative Hubble-induced mass term, 
the AD mechanism becomes viable.


\subsection{Discussion on the origin of the effective operator $K =|\phi|^2|I|^2 /\tilde{M}^2$}
\label{Kefforigin}

Before studying the moduli stabilization, 
let us discuss various possibilities of the origin of 
the effective operator in Eq. (\ref{Keff}).
The structure of this term strongly depends on what the inflaton $I$ is, e.g. its K\"ahler potential
\cite{Dutta:2010sg}. For instance, the volume modulus inflation \cite{Conlon:2008cj} and the warped D-brane inflation 
\cite{Kachru:2003sx}
are not suitable.
In this paper, instead, we
treat the inflaton as the usual chiral matter-like fields including open string moduli
on the visible brane for reheating the visible sector: 
$K_{\rm inflaton} \sim |I|^2$.

When a massive mode is propagating between the inflaton and the AD field,
the cutoff $\tilde{M}$ in the K\"ahler potential (\ref{Keff}) is given by the mass
after integrating it out.
In the string theories, there are many mass scales such as  Kaluza-Klein (KK) scale,
the string scale and winding scale.
With the 6D compactification volume of extra dimension ${\cal V} \equiv R^6M_{\rm string}^6$,
they are written as\footnote{
Here the string scale comes from the coefficient in front of the 10D Einstein term:
$M_{\rm string}^8\int *_{10}{\cal R} \rightarrow M_{\rm string}^2{\cal V}\int *_4 {\cal R} \equiv
M_{\rm pl}^2\int *_4 {\cal R}$ up to the string coupling. The winding modes here are 
open strings
which are wrapping on a cycle in the extra dimension or which are stretching between two branes.
}
\begin{align}
M_{KK} \sim \frac{1}{R} \sim \frac{M_{\rm pl}}{{\cal V}^{2/3}} , ~~~
M_{\rm string} \sim \frac{M_{\rm pl}}{{\cal V}^{1/2}} ,~~~
M_{\rm wind} \sim M_{\rm string}^2 R \sim \frac{M_{\rm pl}}{{\cal V}^{1/3}} 
\end{align}
in the large volume limit. Here we ignored the string coupling dependence.
Note that the above KK and winding modes are the ones propagating in the bulk 
while the local KK and winding modes relevant to the local models
are estimated as
$M_{KK}' \sim M_{\rm pl}/({\cal V}^{1/2}\tau^{1/4})$ and $M_{\rm wind}' \sim M_{\rm pl}\tau^{1/4}/{\cal V}^{1/2}$
respectively, where $\tau$ is a local modulus describing a 4-cycle.
There will be also massive modes in the Landau level on the magnetized branes
and its T-dual modes among intersecting D-branes, and their masses are given by
of order $M_{KK}~ ({\rm or}~M_{KK}')$ and of order $M_{\rm string}$, respectively.
If such massive modes are coupled to the light modes of our interest, 
the operator in Eq. (\ref{Keff}), where $\tilde{M}$ is identified with $M_{KK}, M_{\rm string}$, $M_{\rm wind}$ etc.,
will appear after integrating them out~\cite{Choi:2006bh}.
Then the enhancement factor of the Hubble-induced mass term reads
\begin{align}
c \sim {\cal V}^{2/3}, \qquad {\cal V}^{1/2}  ~~~~ {\rm or}~~~~ {\cal V}^{1/3}, 
\end{align}
with the canonically normalized fields.
Thus it is possible to realize
\begin{align}
c \gg 1
\end{align}
in the large ${\cal V}$ limit. Note that the condition (\ref{condAterm}) is not satisfied for the above values of $c$,
and so, the phase of the AD field starts to oscillate before the radial component.

\subsubsection{Field-theoretical interpretation}

Here let us interpret the behavior discussed above from the view point of the field theory.
We emphasize again that after integrating out massive mode propagating between the light modes,
the cutoff $\tilde{M}$ is given by the mass of the massive mode. 
The low-energy effective theories must be such that the cutoff 
$\tilde{M}$ corresponds to the scale of the massive mode.

Consider first the effective theory with heavy chiral multiplets $(\Phi_H + \Phi_H^c)$\footnote{
On the flat space in the extra dimension, $\Phi_H$ can not be KK mode because the momentum conservation is violated
while massive modes in the Landau level on the magnetized brane (and its T-dual) would be viable even in that case.}
: 
\begin{align}
K &= Z_H (|\Phi_H|^2 + |\Phi_H^c|^2) +Z_I |I|^2 + Z_{\phi} |\phi|^2 , \\
W_{\rm heavy} &= M_H \Phi_H \Phi^c_H + \Phi_H^c \phi I.    
\label{HSuper}
\end{align}
Here $M_H =M_{\rm pl}$ is expected from the holomorphicity\footnote{
If $\Phi_H \Phi_H^c$ is charged under an anomalous $U(1)$ symmetry, 
$M_H$ will be a function of the moduli; it is possible to obtain 
$M_H \ll M_{\rm pl}$
at the non-perturbative level.
We ignored possible term of $K \supset \Phi_H^{\dag}I\phi$ for a simplicity.
} 
and $Z$ is the K\"ahler metric that depends on closed string moduli.
After integrating out $(\Phi_H + \Phi_H^c)$ 
and the canonical normalizing $I_c = Z_I^{1/2}I$ and $\phi_c = Z_\phi^{1/2}\phi$,
we obtain the effective operator, 
\begin{align}
K_{\rm low} = \frac{|I_c|^2 |\phi_c|^2}{\tilde{M}^2} ,~~~{\rm with}~ 
\tilde{M}^2 =  \frac{Z_\phi Z_I}{Z_H}M_{\rm pl}^2.
\end{align}
The $\tilde{M}$ in string theories should correspond to the mass scale discussed above, 
even though it is difficult to obtain the information of the relevant wavefunction $Z$. 

In global string models sitting on the bulk, 
$\tilde M$ will be of order the Kaluza-Klein scale.
In local models, it will be around the string scale for the cancelled RR-tadpole case
or a winding mode scale for the (global) uncancelled RR-tadpole case \cite{Conlon:2009qa}.
Those are because there are no Kaluza-Klein modes on the local brane 
and for the latter case the winding modes are necessary to cancel the global RR-tadpole.

Next, let us consider the case with a heavy $U(1)$ vector multiplet $V_H$:
\begin{align}
K = |I_c|^2 e^{2 g q_I V_H} + |\phi_c|^2 e^{2 g q_\phi V_H} + \frac{1}{2}M_H^2 V_H^2.
\label{HKahler}
\end{align}
Here $g$ is the $U(1)$ coupling, $q_I$ and $q_\phi$ are the $U(1)$ charges of $I$ and $\phi$.
Since we have assumed that inflaton $I$ is a matter-like field, it is plausible for $I$ to have the $U(1)$ charge.
Let us consider the origin of the mass term in two ways.
In string theories, $V_H$ is typically a (non-)anomalous $U(1)$ vector multiplet and 
hence the mass term $M_H$ comes from the St\"uckelberg coupling between the $U(1)$ and string moduli $\Phi$
\begin{align}
K_{\rm moduli} = K(\Phi + \Phi^{\dag} + V_H) \supset K' V + \frac{1}{2} K'' V_H^2  \equiv 
\xi_{\rm FI} V + \frac{1}{2g^2}M_H^2 V_H^2 .
\label{HModuli}
\end{align}
Then $M_H$ is given by the K\"ahler metric of the relevant moduli in the vanishing Fayet-Iliopoulos (FI) term limit, 
$\xi_{\rm FI}=0$.
On the other hand, if the above FI term is non-zero, i.e. $K' \neq 0$, a canonically normalized matter-like field $\psi_c$ charged 
under the $U(1)$ symmetry
will condense  to cancel it via the D-term condition, $D \sim K' + q_{\psi}|\psi_c|^2 \sim 0$.
Then one will find the mass term of $V_H$:
\begin{align}
K_{\psi}= |\psi_c|^2e^{2gq_{\psi}V_H} ~~~ \to ~~~
M_H^2 \sim g^2 \langle |\psi_c|^2 \rangle \sim g^2 |K'|,
\end{align}
for $|K'| \gg K''$.
After integrating out $V_H$ via $\partial_{V_H} K = 0$ in Eq. (\ref{HKahler}) up to the kinetic term of the gauge field, 
one finds 
\begin{align}
K_{\rm low} &= -4g^2 q_I q_\phi \frac{|I_c|^2 |\phi_c|^2}{M_H^2} ; \\
\tilde{M}^2 &= \frac{M_H^2}{4 g^2 |q_\phi q_I|} = \frac{K''}{4  |q_\phi q_I|} ~~~{\rm or} ~~~\frac{|K'|}{4  |q_\phi q_I|}
 ~~~{\rm for}~ q_I q_\phi < 0.
\end{align}
Note that the Goldstone multiplet eaten by $V_H$ is a modulus $\Phi$ in the former case, 
whereas it is $\psi_c$ in the latter case.
Here $M_H$ will be of order string scale for anomalous $U(1)$ case, 
whereas for the case of non-anomalous massive multiplet \cite{Antoniadis:1999ge}
it will be less than of order 
the Kaluza-Klein scale and the relevant gauge coupling will depend on compactification volume.
This is because the non-anomalous $U(1)$ multiplet becomes massive 
due to the anomaly in the compact extra dimension.
We also provide a few examples in the appendix.


\section{LARGE volume scenario with three K\"ahler moduli in the geometric regime}
\label{sec:3}

In this section, we shall study a supergravity model inspired by the string theory, in
which there are lots of string moduli through the compactification.
In a string model compactified on a Calabi-Yau space, 
there will be the dilaton $S$ determining the string coupling,  
the K\"ahler (volume) moduli $T$
and the complex structure (shape) ones $U$.
They should be stabilized because an ultralight moduli can mediate fifth forth among matter fields
and the moduli determine the size of not only compactification but also of physical parameters such as a gauge coupling.
Therefore moduli stabilization is mandatory in string theories.
For it, closed string flux backgrounds in extra dimensions, i.e. flux compactifications, 
are powerful tools to fix a lot of the moduli simultaneously.
The remaining moduli which is not stabilized by the fluxes can become massive by instantons/gaugino condensation.
By combination of them, all the moduli can be stabilized on a Calabi-Yau space \cite{Kachru:2003aw}.
In the followings, we will use the language in the type IIB orientifold flux compactification 
on the Calabi-Yau space \cite{Blumenhagen:2006ci}.

On the other hand, the origin of the low cutoff scale ${\tilde M}$ in Eq. (\ref{Keff}) 
will be naturally explained in terms of
large volume compactification.
However, such moduli relevant to the large volume typically has a long lifetime 
and hence may cause cosmological disasters,
diluting any pre-existing baryon asymmetry.
In the following subsections, the cutoff scale and the dilution factor will be estimated.
The supergravity computations will be done in the Planck unit ($M_{\rm pl} =1$), 
whereas the Planck scale will be shown explicitly in the physical quantities.

\subsection{Moduli stabilization and its consequences}

\subsubsection{Moduli sector}

From this subsection, let us study an explicit supergravity model, 
which is so-called LARGE volume scenario (LVS) \cite{Balasubramanian:2005zx}
studied in \cite{Blumenhagen:2008kq}. (See also \cite{Conlon:2005ki} and \cite{Cicoli:2008va}.)
In such a scenario, a swiss-cheese type Calabi-Yau geometry and a negative Euler number on it\footnote{
With a singular cycle and a proper quantum effect from the large volume cycle, 
a positive Euler number will be also available for LVS \cite{Cicoli:2012fh}.
} will be required.

In this type IIB orientifold model, the dilaton and the complex structures will be stabilized by the fluxes while the
K\"ahler moduli will be fixed by the non-perturbative effect.
We will focus on the K\"ahler moduli $T$ relevant to the low energy physics
and then mention the expected consequences of the dilaton and the complex structure moduli.
The effective theory of K\"ahler moduli $T_i~(i=1,2,3)$ 
are given by
\begin{align}
K &= -2 \log\bigg({\cal V} + \frac{\xi}{2}\bigg), \\
{\cal V} &= (\eta_1\tau_1)^{3/2} - (\eta_2\tau_2)^{3/2} -(\eta_3\tau_3)^{3/2} ,
\qquad \xi = -\frac{\chi \zeta(3)}{2(2\pi)^3 g_s^{3/2}} ,\\ 
W & = Ae^{-a(T_2 +C_1 e^{-2\pi T_3})} -B e^{-b(T_2 +C_2 e^{-2\pi T_3})}, \\
& = Ae^{-aT_2} -B e^{-bT_2} -(aAC_1e^{-aT_2} - bBC_2 e^{-bT_2}) e^{-2\pi T_3} + \cdots ,\\
& \equiv W_{\rm eff} + A_{\rm eff}e^{-2\pi T_3} + \cdots .
\end{align}
In the followings, we will take
\begin{align}
\eta_{i=1,2,3}  = {\cal O}(0.1 - 1) ,  \quad g_s = {\cal O}(0.1), 
\quad  C_{i=1,2},~a,b ={\cal O}(1),  \quad
\chi = - {\cal O}(100) < 0,  \\
A, B  \gg 1 . 
~~~~~~~~~~~~~~~~~~~~~~~~~~~~~~~~~~~~~~~~~
\end{align}
Although choosing $A,B \gg 1$,
$\langle W_{\rm eff} \rangle \ll 1$ will be obtained naturally due to 
the exponentially suppressed racetrack stabilization.
Such large parameters may be explained through fluxed branes.
We decomposed moduli fields as
\begin{align}
T_i & = \tau_i + i \sigma_i .
\end{align}
It is easy to note that $\tau_1$ becomes the lightest modulus and $\sigma_1$ 
is massless because the former is not included in the superpotential 
and the latter is absent from the whole potential.
Here $g_s$ is the string coupling fixed via flux compactification and 
$\chi$ is the Euler number of the Calabi-Yau space: $\chi = 2(h^{11}-h^{21})= 2(3-h^{21}) < 0$. 
(Here we assumed $h_{-}^{11} = h_{+}^{21} =0$.)
$\tau_1$ originates from the bulk volume 4-cycle of the extra dimension while
$\tau_{2}$ and $\tau_3$ come from the local 4-cycle volume. Their axionic partners $\sigma_i$ 
do from RR 4-form on the relevant 4-cycles.
Three $\eta_i$ are related with the intersection number in the Calabi-Yau space between three 2-cycles $\sim \tau_i^{1/2}$.
The superpotential comes from the double gaugino condensations on the separate two stacks of D7-branes wrapping 
on the rigid 4-cycle (divisor)\footnote{
Here, on the D-brane wrapping on the rigid cycle, 
there are no light adjoint matter fields called open string moduli which are Wilson line and the brane position.
A stack of D7-branes sitting on the divisor preserves the supersymmetry and,
in this paper, we will use 4-cycle as the same meaning of divisor.
}
whose volume is given by $\tau_2$, and hence one will find that $a = 2\pi/N$ and $b=2\pi/M$, in which $N$ and $M$ are the rank 
of the gauge group, are of order unity.
Hence the relevant holomorphic gauge coupling $f$ on the branes are given by $\tau_2$, and is supposed to be 
corrected by the Euclidean D3-instanton effects wrapping on non-rigid 4-cycle, whose volume is $\tau_3$: 
$f_i = T_2 + C_i e^{-2\pi T_3}+$ dilaton. Non-perturbative effect with such a gauge coupling
$e^{-f_i}$ is called poly-instanton \cite{Blumenhagen:2008ji}\footnote{
In general, there would be a case that an instanton wrapping on the non-rigid 4-cycle
would contribute not to the superpotential directly but to the gauge coupling on a brane wrapping on the rigid cycle,
because of too many fermionic zero modes on the non-rigid brane. In such a case,
the modulus  relevant to the non-rigid cycle 
would be stabilized via poly-instanton, i.e. gaugino condensations/instantons, 
generated on the rigid brane. (However, K3 divisor is available only if there are some additional mechanisms on it, e.g. 
three form fluxes 
\cite{Blumenhagen:2012kz}.)
Furthermore, this effect becomes important for generating larger mass scale hierarchy between moduli
rather than those of usual instantons \cite{Cicoli:2011yy}.
}. 
Note that we set $\langle W_{\rm flux} \rangle = \langle \int G_3 \wedge \Omega \rangle = 0$ 
and it seems that the number of such flux vacua is not too small \cite{DeWolfe:2004ns}\footnote{
In the literature, it was found that $N_{\rm vacua}(W_{\rm flux}=0)/N_{\rm vacua} \sim 1/L^{{\cal D}/2}$, where
$L$ is the upper limit of flux quanta and ${\cal D}$ is an integer.
Cases with $L={\cal O}(10 -10^3)$ (depending on the RR-tadpole cancellation) and ${\cal D} ={\cal O}(1)$ were considered there.
}. Here $G_3$ is the three-form flux and $\Omega$ is holomorphic three-form on the relevant Calabi-Yau space.

The supergravity potential is written by
\begin{align}
V_F = e^K \bigg[
K^{I\bar{J}}D_I W (D_J W)^{\dag} -3|W|^2
\bigg], ~~~{\rm where}~D_IW = \partial_I W + W(\partial_I K) .
\end{align}
Here $K^{I\bar{J}}$ is the inverse of the K\"ahler metric $K_{I\bar{J}}= \partial_I \partial_{\bar{J}}K$.

One can easily check the vacuum structure\footnote{
We numerically checked the followings are correct.
}.
First we define the gravitino mass 
\begin{align}
m_{3/2} = e^{K/2}W \sim \frac{\langle W_{\rm eff} \rangle}{{\cal V}}
\end{align}
and $F$-terms of SUSY-breaking order parameters are given by
\begin{align}
F^I = -e^{K/2}K^{I\bar{J}}(D_{J} W)^{\dag} .
\end{align}
The heaviest modulus $T_2$ is stabilized near supersymmetric location like the racetrack model\footnote{
See, for instance, \cite{Higaki:2011me} for an explanation about the F-term structures.
}:
\begin{align}
F^{T_2} \sim \partial_{T_2} W &\sim \partial_{T_2} W_{\rm eff} \sim 0
 ~~~ \rightarrow ~~~ \langle T_2 \rangle \simeq  \frac{1}{a-b} \log \bigg( \frac{aA}{bB}\bigg) , \\
 m_{\tau_2} \simeq m_{\sigma_2} & \sim \langle W_{\rm eff} \rangle \ll 1, \qquad \frac{F^{T_2}}{2\tau_2} \sim m_{3/2} \frac{m_{3/2}}{m_{\tau_2}} .
\end{align}
Next, the $T_3$ will be also fixed near the supersymmetric solution like the minimal LVS \cite{Balasubramanian:2005zx}:
\begin{align}
F^{T_3} &\sim 0
 ~~~ \rightarrow ~~~ {\cal V} \sim  \tau_1^{3/2} 
\sim \frac{W_{\rm eff}}{A_{\rm eff}}e^{2\pi T_3}, \\
 m_{\tau_3} \simeq m_{\sigma_3} & \sim \log({\cal V})m_{3/2} \sim  \log({\cal V}) \frac{\langle W_{\rm eff} \rangle}{{\cal V}}, 
\qquad \frac{F^{T_3}}{2\tau_3} \sim m_{3/2} \frac{m_{3/2}}{m_{\tau_3}}  .
\end{align}
Finally, the lightest modulus $T_1$, which is no-scale modulus corrected 
by the non-perturbative superpotential and $\xi {\cal V}^{-1}$ in the K\"ahler potential, 
will be stabilized because of the $\xi$-dependent term like the minimal LVS case:
\begin{align}
\langle T_3 \rangle \sim \frac{\langle \log(\tau_1^{3/2}) \rangle }{2\pi} \sim \xi^{2/3}, \qquad m_{\tau_1} \sim \frac{\langle W_{\rm eff} \rangle}{\sqrt{\log({\cal V})}{\cal V}^{3/2}}, \qquad
\frac{F^{\tau_1}}{2\tau_1} = m_{3/2} \left ( 1 + {\cal O}(\log({\cal V})^{-1}{\cal V}^{-1}) \right).
\end{align}
Note that $\sigma_1$ is massless axion and modulino $\tilde{T}_1$ is the goldstino.
Thus one obtains ${\cal V} \simeq \tau_1^{3/2} \gg \tau_2^{3/2} \sim \tau_3^{3/2} $. Because of the large volume, even 
if there is a superpotential $\delta W \sim e^{-2\pi T_1}$, $\sigma_1$ still stays almost massless.
As a result, the vacuum is non-supersymmetric AdS: 
\begin{align}
\langle V_F \rangle \sim - \bigg\langle\frac{|W_{\rm eff}|^2}{\log({\cal V}){\cal V}^3}\bigg \rangle < 0.
\end{align}
Following the KKLT proposal \cite{Kachru:2003aw, Choi:2005ge}, 
the uplifting term is required to obtain the tiny cosmological constant $\langle V \rangle \approx 0$, 
where
\begin{align}
V &= V_F + V_{\rm uplift}, \\
V_{\rm uplift} &= \epsilon e^{2K/3} = \frac{\epsilon}{{\cal V}^{4/3}} , ~~~
\epsilon \sim \bigg\langle \frac{|W|^2}{\log({\cal V}){\cal V}^{5/3}} \bigg\rangle \ll 1.  
\end{align}
Here we have assumed that $V_{\rm uplift}$ is generated on the sequestered anti-D3-branes on a top of the warped throat\footnote{
Dynamical SUSY breaking model is also viable \cite{Cicoli:2012fh}. 
See also \cite{Saltman:2004sn} for the KKLT case and discussion on such an anti-brane from the view point of 
the 10D supergravity \cite{Bena:2012bk}.
}
and $\epsilon$ means the minimum of the warp factor.

Let us consider to include the dilaton and complex structure moduli.
They will be stabilized, being consistent with the K\"ahler moduli $T_i$, because
the potential for them is of ${\cal O}({\cal V}^{-2})$ with $W_{\rm flux}$ whereas 
the K\"ahler moduli are done at ${\cal O}({\cal V}^{-1})$ for $T_2$ and at ${\cal O}({\cal V}^{-3})$ 
for $T_1$ and $T_3$. Then, their mass scale will be of order $M_{\rm pl}/{\cal V}$.
In this case, the contribution from the dilaton $S$ and complex structure moduli $U$ is written as
\begin{align}
K_{\rm moduli} = -\log(S+S^{\dag}) +K_T (T+T^{\dag}) + K_U (U+U^{\dag}).
\end{align}
And then
\begin{align}
m_{S,U} \sim \frac{W_{\rm flux}''}{K''} e^{K/2} \sim \frac{1}{{\cal V}}.
\end{align}
Here note that the supersymmetric mass terms between $S$ and $U$ are generally included in $W_{\rm flux}$.
In this section, we assume that $F^U = 0$ and
\begin{align}
\langle F^S \rangle 
\sim \frac{1}{\cal V}((\partial_S {K_{\rm moduli}}) \langle W \rangle + \partial_S W_{\rm flux})  
\sim  \frac{(\partial_S K_T)}{\log({\cal V})\cal V} \langle W_{\rm eff} \rangle
\sim  \frac{\langle W_{\rm eff} \rangle}{\log({\cal V}){\cal V}^2}  .
\end{align}
This is because one expects that $S$ would be stabilized almost without $\xi$-dependent K\"ahler potential: 
$-W/(S+S^{\dag})+ \partial_S W_{\rm flux} \approx 0$, though
its magnitude would depend on flux models strongly \cite{Abe:2006xi, Blumenhagen:2009gk}.
Hence we will not consider also the derivative of $F^S$ with respect to any moduli.

\subsubsection{Visible sector}

Next, let us focus on the visible sector in this local model, and suppose that 
the minimal supersymmetric standard model (MSSM) is localized on the rigid D7-brane wrapping on the $\tau_2$ 4-cycle.
Therefore the relevant effective potential will be given by \cite{Conlon:2006tj}\footnote{
In this case, an instanton on $\tau_3$ would contribute to the visible gauge coupling 
while the gaugino condensations on $\tau_2$ would have effect on the visible small Yukawa coupling.
However, we neglected those effects for a simplicity.
The contribution of an instanton effect $4\pi \delta f_{\rm vis} = e^{-2\pi T_3}$ in the holomorphic gauge coupling
to the gaugino mass will be the same order as computation shown below at most.
For non-perturbative Yukawa coupling, 
even if there would be terms $\delta W_{\rm Yukawa} = e^{-a T_2} \phi^3$ 
or $e^{-a T_2}e^{-2\pi T_3} \phi^3$ from gaugino condensations, 
one will not suffer from the flavor changing neutral current induced by
non-perturbative $A$-term because $m_0$ is large and $F^{T_2}$ is small.}
\begin{align}
K_{\rm vis} = \frac{(T_2 +T_2^{\dag})^{\lambda_i}}{(T_1+T_1^{\dag})}|\phi_i|^2 =Z_i |\phi_{i}|^2 , \qquad
f_{{\rm vis}, a} = \frac{1}{4\pi}( T_2 + h_{a} S ) , \qquad 
W_{\rm vis} = W_{\rm MSSM} (\phi_{\rm vis}).
\end{align}
Here we will take a minimal model $\lambda_i = 1/3$ and assume the approximate
gauge coupling unification, i.e. $h_a ={\cal O}(1)~(a=1,2,3)$,
which is given by the world volume flux depending on the gauge group on the visible brane \cite{Blumenhagen:2008aw}. 
For instance $h_2 < h_3 < h_1$ is possible, depending on the model. Notice that the gauge fluxes on the visible brane 
are relevant for obtaining the chiral matter spectra.
$S = 1/g_s - i C_0^{\rm RR}$, where $C_0^{\rm RR}$ is the RR scalar zero mode, 
is the 4D string axion-dilaton fixed by closed string flux.
The factor $1/(T_1+T_1^{\dag})^{-1}$ is important because the physical couplings in the superpotential 
are (almost) independent on the $T_1$ in the large ${\cal V}$ limit:
\begin{align}
W = y \frac{\phi_{\rm vis}^n}{M_{\rm pl}^{n-3}} ~~~\to ~~~
W_{\rm global} &= e^{K_{\rm moduli}/2}\frac{y}{\sqrt{Z^3}} \frac{\phi_{\rm vis}}{\sqrt{Z^{n-3}}M_{\rm pl}^{n-3}} \\
& = y^{\rm phys}\frac{\phi_{ c}^n}{M_{*}^{n-3}}.
\end{align}
Thus one finds $y^{\rm phys} = y(\tau_2)$ and 
\begin{align}
M_{*} \simeq \frac{M_{\rm pl}}{{\cal V}^{1/3}} \sim M_{\rm wind}.
\end{align}
The cutoff scale in the superpotential would be of order winding mode scale \cite{Conlon:2009qa, Choi:2010gm}.
Therefore we will not consider the moduli-redefinition effect \cite{Conlon:2010ji} in this whole paper.

The SUSY-breaking soft masses at the messenger scale are given by\footnote{We added 
$2\langle V_{\rm uplift}\rangle/3$ contribution to the soft scalar mass from the uplifting potential.}
(see, e.g. \cite{Brignole:1997dp})
\begin{align}
m_0^2 &\simeq  - F^I (F^{\bar{J}})^{\dag}\partial_I \partial_{\bar{J}} \log(e^{-K_{\rm moduli}/3}Z) \\
& \sim \frac{1}{{\cal V}}\bigg|\frac{F^{T_1}}{2\tau_1}\bigg|^2 \simeq \frac{m_{3/2}^2}{{\cal V}} 
\sim \frac{\langle |W_{\rm eff}|^2\rangle}{{\cal V}^3}, \\
M_a & \simeq F^I \partial_I \log(f_{{\rm vis},a})  
+ \frac{\alpha_a}{4\pi} 
\left( b_a F^{\varphi} - 2 \sum_i {\rm tr}(T_a^2(\phi_i))F^I \partial_I \log(e^{-K_{\rm moduli}/3}Z_i)
\right)
\\ 
& \simeq 
h_{a} \frac{F^S}{2\tau_2} 
+ \frac{\alpha_a}{4\pi}\frac{1}{{\cal V}}\frac{F^{T_1}}{2\tau_1} 
 \sim \frac{m_{3/2}}{\log({\cal V}){\cal V}} \sim \frac{\langle W_{\rm eff}\rangle}{\log({\cal V}){\cal V}^2}, \\
A_{i_1 \cdots i_n} & \simeq (n-3) F^{\varphi} -F^I \partial_I \log\bigg(
\frac{y_{i_1 \cdots i_n}}{e^{-nK_{\rm moduli}/3}Z_{i_1} \cdots Z_{i_n}}
\bigg) \\
& \sim  \frac{1}{{\cal V}}\frac{F^{T_1}}{2\tau_1} 
+ \frac{F^S}{S+S^{\dag}}
 \sim \frac{m_{3/2}}{{\cal V}} 
\sim \frac{\langle W_{\rm eff}\rangle}{{\cal V}^2}.
\end{align}
Here the second term in the gaugino mass is originating from the anomaly mediation; 
$b_a~(a=1,2,3)$ is 1-loop coefficient of beta function of the gauge coupling in the MSSM and
${\rm tr}(T_a^2(\phi_i))$ is dynkin index for the matter $\phi_i$ charged under the MSSM gauge group.
Note that the anomaly mediation contribution would be suppressed or be at most the same order compared to $F^S$ 
in the gaugino mass, 
because the F-term of the compensator 
\begin{align}
F^{\varphi}  = m_{3/2} + \frac{\partial_I K }{3}F^I
\sim
\frac{F^S}{S+S^{\dag}}
+ \frac{m_{3/2}}{\log({\cal V}){\cal V}} \sim \frac{\langle W_{\rm eff} \rangle}{\log({\cal V}){\cal V}^2} 
\end{align}
will be also suppressed.

The above results of the soft masses are easily understood: 
Because $F^{T_2}$ is negligible due to the very massive $T_2$,
SUSY-breaking structure in the visible sector is determined by the no-scale modulus $T_1$ 
with ${\cal V}^{-1}$ corrections in the potential.
(However, in the computation of the coupling between the gauginos and moduli, the $F^{T_2}$
is important.)
Note that the perturbative Peccei-Quinn symmetries of moduli/dilaton $T, S \to T, S + {\rm const.}$ 
forbid the moduli/dilaton dependence of couplings $y_{i_1 \cdots i_n}$ in the visible superpotetial.

Finally  the mass spectrum is summarized as
\begin{align}
m_{\tau_2} > m_{S,U} \gg m_{\tau_3} > m_{3/2} > m_0 \gtrsim m_{\tau_1} \gg A_n > M_a .
\end{align}
This is a realization of split SUSY \cite{ArkaniHamed:2004fb}.
What is important is that the decay of the lightest modulus to gravitinos is kinematically forbidden
because of the fact that $m_{3/2} \gg m_{\tau_1}$.
Recall that axion $\sigma_1$ is massless.

\subsubsection{A numerical example based on the model 2 in the literature \cite{Blumenhagen:2008kq}}

Following the literature, we hereafter will use the following parameters for numerical calculations
\begin{align}
\eta_1 &=1, \quad \eta_2 =\frac{1}{53} ,\quad \eta_3 = \frac{1}{6} , \quad \chi = -136, \quad g_s = \frac{2}{5},
 \\
C_1 &=1, \quad C_2 =3, \quad a= \frac{2\pi}{8}, \quad 
b= \frac{2\pi}{9} , \\
A & = 1.28 \times 10^{5} , \quad B = 1.6 \times 10^{4} .
\end{align}
$A$ and $B$ are much larger than unity, however this fact may be explained in terms of gaugino condensation on the magnetized 
D7-brane \cite{Abe:2005rx, Haack:2006cy}: 
\begin{align}
A \sim e^{-a(h_A/g_s )} \gg 1 , \qquad B \sim  e^{-b(h_B/g_s) } \gg 1. 
\end{align}
Here $h_{A,B} \sim \frac{-1}{4\pi^2}
\int_{\Sigma_{A,B}} F_{A,B} \wedge F_{A,B} - {\cal R}_{A,B} \wedge {\cal R}_{A,B} $ 
is assumed to be negative and be of ${\cal O}(1)$. 
$\Sigma_{A,B}$ are respectively 
the 4-cycle worldvolume in the extra dimension of two branes $(A,B)$ which experience gaugino condensations.
$F_{A,B}$ and ${\cal R}_{A,B}$ are magnetic flux and geometric curvature two-form on the relevant two branes.
In this example, in spite of $A, B \gg 1$, one finds 
\begin{align}
\langle W_{\rm eff} \rangle \ll 1 
\end{align}
and hence the scales of gaugino condensations is lower than the cutoff scale.
Notice that $C_i ={\cal O}(1)~(i=1,2)$ will be also 
plausible when the instanton on $\tau_3$ is not fluxed 
and the geometric curvatures relevant to $\tau_3$ are zero.

With these numerical parameters, the vacuum expectation values of moduli are read as
\begin{align}
\langle {\cal V} \rangle = 78559, \qquad 
\langle T_2 \rangle = 25.18 , \qquad 
\langle T_3 \rangle = 2.88 . 
\end{align} 
Other parameters at the vacuum is shown in TABLE \ref{TableI}.
Although $m_{\tau_2}$ exceeds the string scale in this numerical example as shown, 
we are going to continue the discussion;
it is expected that a proper choice of parameters would bring us to a viable region 
in which $m_{\tau_2}$ is below the string scale without changing the essence.
In addition, we do not include complex structure moduli dependence in the scalar potential of $e^K$ in the numerical computation.
This factor would reduce the masses slightly.
At any rate, it is smaller than the Planck scale and hence this model could be valid just as the supergravity model.

\begin{table}[ht]
\begin{center}
\begin{tabular}{|c|c|c|c|c|} \hline 
Fundamental parameters   & Moduli masses & F-terms & Soft masses  \\ \hline \hline
$M_{\rm string}  = 1.2 \times 10^{16}$ GeV & $m_{S,U} = 3.4 \times 10^{13}$ GeV & $F^S = {\cal O}(10^3)$ GeV   
&  $m_{3/2} = 1.4 \times 10^{9}$ GeV  \\
$M_* = 7.9 \times 10^{16}$ GeV & $m_{\tau_2, \sigma_2} = 3.1 \times 10^{17}$ GeV & $F^{T_2}/2\tau_2 = {\cal O}(10)$ GeV 
& $m_0 ={\cal O} (10^6 - 10^7)$ GeV
\\
$|W_{\rm eff}|/M_{\rm pl}^3 = 4.1 \times 10^{-5}$ & $m_{\tau_3, \sigma_3} = 5.3 \times 10^{10}$ GeV  & 
$F^{T_3}/2\tau_3 = 2.9 \times 10^7$ GeV 
&  $A_n = {\cal O} ( 10^3 - 10^{4})$ GeV
\\
$\langle {\cal V} \rangle = 78559$ & $m_{\tau_1} = 1.5 \times 10^{6}$ GeV  &  $F^{T_1}/2\tau_1 = 1.4\times 10^{9}$ GeV
& $M_{1/2} = {\cal O}(10^3)$ GeV  \\ \hline
\end{tabular}
\caption{A numerical example. Here we have included the dilaton dependence $e^{K} \supset 1/2g_s$ in the supergravity potential.
Although $m_{\tau_2}$ is greater than the string scale, it is smaller than the Planck scale;
this model could be valid just as the supergravity model. 
We assumed $F^S \sim F^{\varphi} = O(1-10)$ TeV.
The magnitude of $F^{T_1}$ strongly depends on the accuracy of computation because $T_1$ is so heavy.
Note that $\sigma_1$ is massless axion decoupled from the visible sector.
}
\label{TableI}
\end{center}
\end{table}

\subsubsection{Higgs sector: $\mu/B\mu$-term}

So far we have studied the moduli sector and the visible sector except for Higgs fields $(H_u, H_d)$ in the MSSM.
In the Higgs sector, generation of $\mu/B\mu$-term is important.
Furthermore, because SUSY-breaking stop mass is so heavy as $10^{6}$ GeV in this model,
the Higgs quartic coupling $\lambda_{\rm tree} \simeq (g_2^2 +g_Y^2)\cos^2(2\beta)/8$ at a high scale 
should be almost vanishing to obtain $m_{\rm Higgs} \simeq 125$ GeV for the lightest Higgs boson.
In other words, $\tan \beta ={\cal O}(1)$ is required \cite{Ibe:2011aa}.
This can be naturally explained if the approximate shift symmetry exists in the Higgs fields \cite{Hebecker:2012qp}.

If $(H_u, H_d)$ have the origin from a higher dimensional gauge field (SUSY gauge-Higgs unification) 
through the gauge symmetry breaking, e.g. by boundary condition,
one would find an approximate shift symmetry $H_{u,d} \to H_{u,d} + i \gamma$, which comes from the gauge symmetry.
Here $\gamma$ is the constant.
This constrains the structure of their leading K\"ahler potential at the tree level,
though this symmetry will be broken at the quantum level
or by the coupling to matter fields at the tree level (including worldsheet instantons).
As a result the K\"ahler potential will be then given by
\begin{align}
K_{\rm Higgs} = \frac{(T_2 +T_2^{\dag})^{1/3}}{(T_1 +T_1^{\dag})} |H_u + H_d^{\dag}|^2 + \cdots .
\end{align}
We have omitted terms which are breaking such shift symmetry, and 
it can be expected that they are threshold correction from the massive modes, e.g. KK modes.
What is important is that the Giudice-Masiero term is included:
\begin{align}
K_{\rm Higgs} \supset \frac{(T_2 +T_2^{\dag})^{1/3}}{(T_1 +T_1^{\dag})} H_u H_d .
\end{align}

From this K\"ahler potential, the proper $\mu$-term and $B\mu$-term are obtained
\begin{align}
\mu &\simeq 
m_{3/2} - \frac{F^{T_1}}{2\tau_1}
\sim \frac{m_{3/2}}{\log({\cal V}){\cal V}}
\sim \frac{\langle W_{\rm eff}\rangle}{\log({\cal V}){\cal V}^2} = {\cal O}(1)~{\rm TeV}, \\
B\mu &\simeq |\mu|^2 + m_0^2 = {\cal O}(10^{12})~{\rm GeV}^2.
\end{align}
Here we have added $2\langle V_{\rm uplift}\rangle/3$ contribution to $B\mu$-term and used values in the numerical example.
Although we ignored $F^{T_2}$ in the soft masses, this becomes important for computation of the moduli coupling.
From the extremum condition of the Higgs potential at the electroweak scale, one will find 
\begin{align}
\sin (2\beta) = \frac{2 B\mu}{m_{Hu}^2 + m_{Hd}^2 + 2|\mu|^2} = {\cal O}(1).
\end{align}
It is expected the correction term will come from quantum effects and the threshold correction mentioned above.
Thus $\tan \beta = {\cal O}(1)$ will be realized.
Note that a fine-tuning is now demanded for achieving the correct electroweak symmetry breaking.

\subsection{Moduli problem in LVS}

Here let us discuss the moduli problem, focusing on the lightest modulus.
The purpose of this subsection is to calculate the dilution factor by the lightest modulus decay.

Before uplifting the scalar potential, 
the AdS minimum has the depth of order 
$-|\langle W_{\rm eff} \rangle|^2\log({\cal V})^{-1}{\cal V}^{-3} \simeq -m_{\tau_1}^2 M_{\rm pl}^2$.
The height of the potential barrier is comparable to the depth.
Hence, in order to avoid run-away and decompactification, 
the Hubble parameter during the inflation is constrained as \cite{Conlon:2008cj}
\begin{align}
H_{\rm inf} 
\lesssim m_{\tau_1} .
\label{hinf_const}
\end{align}
This is because the inflaton potential energy $3H_{\rm inf}^2M_{\rm pl}^2$ should be smaller than 
the height of the barrier $m_{\tau_1}^2 M_{\rm pl}^2$ (see also \cite{Kallosh:2004yh})\footnote{
The constraint may be modified, 
if the position or the potential of moduli is changed drastically during the inflation.
For instance, one can consider cases that
a field shifted by the inflaton VEV makes the higher barrier 
or inflaton is the lightest rolling modulus \cite{Conlon:2008cj},
realizing $m_{\tau_1} <  H_{\rm inf}$.
(See also the modulus inflation model realizing $m_{3/2} < H_{\rm inf}$ for non-LVS case \cite{Badziak:2008yg}.)
Hence, we expect that coherent oscillation of moduli with a large amplitude could occur in LVS,
depending on the inflation model for realizing $c^2 >0$.
}.
For an inflation model with $H_{\rm inf} \lesssim m_{\tau_1}$, 
the heavier moduli will stay at the true minima  during inflation,
while the lightest one  may be deviated from the true minimum.
The shift of the modulus VEV during inflation is expected to be of order
\begin{align}
\frac{H_{\rm inf}^2}{m_{\tau_1}^2} M_{\rm pl}
\end{align}
in terms of the canonically normalized modulus. Note however that it does not necessarily mean that
the lightest modulus starts to oscillate with an amplitude given above.
If the upper bound on $H_{\rm inf}$ is saturated, i.e.,  $H_{\rm inf} \sim m_{\tau_1}$, there is a cosmological moduli problem
as usual; the modulus $\tau_1$ starts to oscillate with an amplitude of order the Planck scale, and dominates the Universe
soon after the reheating. The situation is a bit more complicated when  $H_{\rm inf} < m_{\tau_1}$.
At the end of inflation, the inflaton begins to oscillate 
and the potential energy of the inflaton is transferred to the kinetic energy.
If the time scale of the inflation oscillation  is shorter than that of 
the modulus, i.e. $m_{I} \gg m_{\tau_1}$, 
where $m_I$ is the inflaton mass around true vacuum, 
the modulus cannot follow  the change of the potential caused by the inflaton and 
it starts to oscillate around the true vacuum soon after the inflation ends~\cite{Nakayama:2011wqa}.

Thus,  the oscillation amplitude of the modulus is expected to be of order
$\frac{H_{\rm inf}^2}{m_{\tau_1}^2} M_{\rm pl} < M_{\rm pl}$,
when the whole potential energy of inflaton is transmitted into the kinetic energy.
Noting that the inflation scale is also bounded below for the AD mechanism to work,
$H_{\rm inf} > m_0/c$, the resultant moduli abundance can be sizable.
In particular, for the numerical example shown in TABLE \ref{TableI}, 
the inflation scale is
almost comparable to $m_{\tau_1}$, and so, there is a serious
cosmological moduli problem.

On the other hand, if all the time scale of the inflaton dynamics is longer than
that of the modulus, i.e. $m_{I},~H_{\rm inf} < m_{\tau_1}$,
the modulus would settle down to the true vacuum without oscillation
and there will  be no moduli problem.
However, the reheating temperature by inflaton decay  tends to be low for such low-scale inflation models,
and so, the AD mechanism will be an important possibility at any rate.

To summarize, as long as the inflaton mass at the potential minimum is heavier than
the modulus mass, the modulus starts coherent oscillations soon after inflation. Although
the oscillation amplitude is suppressed for a lower inflation scale, it is bounded below
for the successful AD baryogenesis. Therefore, the low-energy theory suffers from the cosmological
moduli problem. In the following we consider the cosmology of the modulus, assuming that
it dominates the energy density of the Universe.

Here let us express the moduli fields in terms of the 
canonically normalized mass eigenstates $\{ \delta \phi ,~ \delta a \}$\footnote{
We have checked this relation numerically. This is consistent with the results studied 
in \cite{Conlon:2007gk} and \cite{Cicoli:2010ha}. 
}:
\begin{align}
\begin{pmatrix}
\delta \tau_1 \\
\delta \tau_2 \\
\delta \tau_3 \\
\delta (1/g_s) 
\end{pmatrix}
\sim 
\begin{pmatrix}
{\cal V}^{2/3} \\
\frac{m_{3/2}}{m_{\tau_2}} \\
\frac{m_{3/2}}{m_{\tau_3}} \\
{\cal V}^{-1/2}
\end{pmatrix}
\delta \phi_1 +
\begin{pmatrix}
{\cal V}^{1/6} \\
{\cal V}^{1/2} \\
{\cal V}^{-1/2} \\
{\cal V}^{-1}
\end{pmatrix}
\delta \phi_2 +
\begin{pmatrix}
{\cal V}^{1/6} \\
{\cal V}^{-1/2} \\
{\cal V}^{1/2} \\
{\cal V}^{-1}
\end{pmatrix}
\delta \phi_3 +
\begin{pmatrix}
{\cal V}^{1/6} \\
{\cal V}^{-1/2} \\
{\cal V}^{-1/2} \\
{\cal O}(1)
\end{pmatrix}
\delta \phi_s , \\
\begin{pmatrix}
\delta \sigma_1 \\
\delta \sigma_2 \\
\delta \sigma_3 \\
\delta C_0^{\rm RR} 
\end{pmatrix}
\sim 
\begin{pmatrix}
{\cal V}^{2/3} \\
0 \\
0 \\
0
\end{pmatrix}
\delta a_1 +
\begin{pmatrix}
{\cal V}^{1/6} \\
{\cal V}^{1/2} \\
{\cal V}^{-1/2} \\
{\cal V}^{-1}
\end{pmatrix}
\delta a_2 +
\begin{pmatrix}
{\cal V}^{1/6} \\
{\cal V}^{-1/2} \\
{\cal V}^{1/2} \\
{\cal V}^{-1}
\end{pmatrix}
\delta a_3 +
\begin{pmatrix}
0 \\
{\cal V}^{-1/2} \\
{\cal V}^{-1/2} \\
{\cal O}(1)
\end{pmatrix}
\delta a_s .
\end{align}
The mass eigenstates $\delta \phi_i$ should not be confused with the AD field $\phi$.
Note that $m_{3/2}/m_{\tau_2} \sim 1/{\cal V}$ and $m_{3/2}/m_{\tau_3} \sim 1/\log({\cal V})$.
Here $(\delta \phi_{2}, \delta a_{2})$, $(\delta \phi_{3}, \delta a_{3})$, 
$(\delta \phi_{s}, \delta a_{s})$ and $\delta \phi_1$ have mass eigenvalues of 
$m_{\tau_2}$, $m_{\tau_3}$, $m_S$ and $m_{\tau_1}$,  respectively,  while
$\delta a_1$ is massless.

Let us estimate the lifetime of $\phi_1$.
The relevant moduli couplings to the visible sector arise from the gaugino mass and $\mu$-term:
\begin{align}
 \lambda_c \lambda_c \bigg[ \frac{F^{T_2}}{2\tau_2} +
\mbox{anomaly-mediated terms} 
\bigg] + \frac{1}{3}\frac{F^{T_2}}{2\tau_2}\tilde h_c \tilde h_c, 
\end{align}
where $\lambda_c$ and $\tilde{h}_c$ are canonically normalized gaugino and higgsino.
The derivative of $F^{T_2}$ with respect to the moduli are given by
\begin{align}
\begin{pmatrix}
\partial_{T_1} \\
\partial_{T_2} \\
\partial_{T_3} \\
\end{pmatrix}
\frac{F^{T_2}}{2\tau_2}
\sim 
\begin{pmatrix}
\frac{m_{3/2}}{{\cal V}^{2/3}} \\
m_{\tau_2} \\
 m_{\tau_3} \\
\end{pmatrix}
.
\end{align}
Hence there is a large mixing between $T_1$ and $T_2$ via $F$-terms, unlike in the singular regime we shall study later.
From this, one can read the interactions relevant to the decay  of the $\delta \phi_1$:
\begin{align}
\delta \phi_1 \frac{m_{3/2}}{M_{\rm pl}}
\left[
\lambda_c \lambda_c  + \tilde{h}_c\tilde{h}_c
\right].
\end{align}
Due to the loop factor,
the anomaly mediation contribution to the gaugino mass is irrelevant 
for  the coupling between the modulus and gauginos.
Thus the modulus decays into gauginos and higgsinos if it is allowed kinematically, and
the decay width to the visible sector  is given by\footnote{
We will use the dimensionless $W_{\rm eff}$ through the normalization of $W_{\rm eff} \to M_{\rm pl}^3 W_{\rm eff}$. 
}
\begin{align}
\Gamma_{\phi_1 }  &\simeq \Gamma({\phi_1} \to \tilde g \tilde g ) + \Gamma({\phi_1} \to \tilde h \tilde h ) \\
&\sim 
\frac{N_{c}}{4\pi} \left(\frac{m_{3/2}}{M_{\rm pl}}\right)^2 m_{\tau_1} 
 \sim
\frac{{\cal V} \log({\cal V})}{4\pi} \frac{m_{\tau_1}^3}{M_{\rm pl}^2}  
\sim \frac{\langle W_{\rm eff} \rangle^3}{\sqrt{\log({\cal V})}{\cal V}^{7/2}} M_{\rm pl} .
\end{align}
Here $N_c (=12)$ is the effective number of decay channel.
Note that the decay of the lightest modulus to gravitinos is kinematically forbidden.
Here, the decay of $\phi_1 \to 2 a_1$ also proceeds via the kinetic term
\begin{align}
\frac{\delta \phi_1}{M_{\rm pl}} (\partial \delta a_1)^2.
\end{align}
The branching fraction of the decay to a pair of axions is estimated as
\begin{align}
B(\phi_1 \to a_1 a_1) \simeq \bigg(\frac{m_{\tau_1}}{m_{3/2}}\bigg)^2 
\sim \frac{1}{\log({\cal V}) {\cal V}} \sim 10^{-6} .
\end{align}
Thus the produced axions from the modulus decay is too small to affect the BBN or CMB.\footnote{
\label{footnoteDR}
The present observations give a slight preference to the existence of dark radiation. The non-thermal production
of axions is an interesting possibility~\cite{Ichikawa:2007jv}.
}

The decay temperature of $\delta \phi_1$ 
in the radiation dominated Universe is estimated as
\begin{align}
T_{\rm dec}^{\phi_1} 
&\simeq \sqrt{\Gamma_{\phi_1} M_{\rm pl}} 
\simeq 1.1 \times 10^3 {\rm GeV} 
\bigg(
\frac{m_{\tau_1}}{1.5 \times 10^6 {\rm GeV}}
\bigg)^{1/2} \bigg(
\frac{m_{3/2}}{1.4 \times 10^9 {\rm GeV}}
\bigg) .
\end{align}
Thus, the lightest modulus $\phi_1$ decays well before the BBN.

The dilution factor due to the modulus decay can be estimated as follows. 
For most of the parameter region of our interest, the $\delta \phi_1$ starts to oscillate soon after inflation.
Hence the temperature when the energy density of $\delta \phi_1$ dominates  the Universe is given by
\begin{align}
T_{\rm dom}^{\phi_1} \simeq 
\frac{T_R}{3} \left(\frac{m_{\tau_1}}{H_{\rm inf}} \right)^2 
\bigg(\frac{\Delta \phi_1}{M_{\rm pl}}\bigg)^2
\simeq \frac{T_R}{3} 
\bigg(\frac{\Delta \phi_1}{M_{\rm pl}}\bigg) .
\end{align}
$\Delta \phi_1$ is the oscillation amplitude of the lightest modulus around the true minimum:
$\Delta \phi_1/M_{\rm pl} \sim H_{\rm inf}^2/m_{\tau_1}^2$ as mentioned above.
Thus the dilution factor by the $\delta \phi_1$ decay is given by
\begin{align}
\Delta^{-1} &\simeq \frac{T_{\rm dec}^{\phi_1}}{T_{\rm dom}^{\phi_1}} 
 \simeq  3.4 \times 10^{-4} 
\bigg(
\frac{m_{\tau_1}}{1.5 \times 10^6 {\rm GeV}}
\bigg)^{1/2} \bigg(
\frac{m_{3/2}}{1.4 \times 10^9 {\rm GeV}}
\bigg) 
\bigg(
\frac{T_R}{ 10^7 {\rm GeV}}
\bigg)^{-1} 
\bigg(\frac{\Delta \phi_1}{M_{\rm pl}}\bigg)^{-1} .
\end{align}

Finally we briefly comment on the neutralino dark matter non-thermally produced  by the decay of the
lightest modulus. We assume that the  LSP is Wino or higgsino-like neutralino (or their combination).
Although a large number of the neutralino LSP is produced by the modulus decay, 
they enter  the thermal bath soon after they are produced,
 because the decay temperature is comparable to the Wino/higgsino mass.
Their abundance is fixed when the pair annihilation rate  becomes smaller than the
expansion rate of the Universe.
After that, they decoupled from the thermal bath and their number  in a comoving volume is fixed.
In the case of the Wino-like neutralino LSP, the thermal relic density  is given by \cite{Hisano:2006nn, Giudice:1998xp}
\begin{align}
\Omega_{\chi}^{(\rm thermal)} h^2 \simeq 0.1  \times
\bigg(
\frac{m_{\chi}}{2.8 \times 10^3 {\rm GeV}}
\bigg)^2 .
\end{align}
If the LSP is the higgsino\footnote{
If $F^S \sim 1/{\cal V}^2$ were assumed, one would find this result similarly. 
},
it is given by~\cite{Hisano:2006nn, ArkaniHamed:2004fb} 
\begin{align}
\Omega_{\chi}^{\rm (thermal)} h^2 \sim 0.1 
 \times 
\left(
\frac{\mu}{10^3{\rm GeV}}
\right)^2 .
\end{align}
Thus the dark matter abundance can be explained by the thermal relic of the Wino- or higgsino-like
neutralino LSP.

\subsection{AD baryogenesis in LVS model in the geometric regime}

In this subsection, we apply this supergravity model to the AD baryogenesis,
assuming that the inflaton is coming from a stack of the visible brane.
There are several new ingredients for the AD baryogenesis,  such as
an enhanced coupling with the inflaton
and the dilution factor $\Delta^{-1}$ by the lightest modulus decay.
Note that the other heavy moduli are irrelevant for the present discussion. 

We assume  that there somehow exists the enhanced coupling with the inflaton in
the K\"ahler potential and a non-renormalizable interaction in the superpotential:
\begin{align}
K= c^2({\cal V}) \frac{|I_c|^2|\phi_c|^2}{M_{\rm pl}^2}, \qquad
W=y  \frac{\phi^n}{M_{\rm pl}^{n-3}} ,~~~{\rm where}~~\phi_c \equiv Z_{\phi}^{1/2}\phi .
\end{align}
We would like to emphasize here that, even if $H_{\rm inf} \lesssim m_{\tau_1} \sim m_0$,
$m_{0} \ll cH_{\rm inf}$ is possible due to the large $c$, and the condition (\ref{conditionforAD}) can be
satisfied. 

While the lightest modulus starts to oscillate soon after inflation, 
the AD field will remain away from the origin due to the inflaton coupling.
Then, effective mass matrix of AD field and the normalized lightest modulus is given by
\begin{align}
\begin{pmatrix}
c^2 H^2 & - \frac{m_{3/2}^2}{{\cal V}} \frac{{\langle \phi_c \rangle}}{M_{\rm pl}}
 + c^2 H^2 \left(\frac{\langle \phi_c \rangle}{M_{\rm pl}} \right) \\
- \frac{m_{3/2}^2}{{\cal V}} \frac{{\langle \phi_c \rangle}}{M_{\rm pl}}
 + c^2 H^2 \left(\frac{\langle \phi_c \rangle}{M_{\rm pl}} \right)
 & m_{\tau_1}^2 
\end{pmatrix}
\label{Mass1}
\end{align}
for $m_0/c < H \lesssim H_{\rm inf}$. (Note that the kinetic term is diagonal up to $\phi_c/M_{\rm pl}$.)
Here we could find contribution to the mixing from 
$\delta \tau_i \delta \phi_c (\partial_{\tau_i} m_0^2) \phi_c \sim 
- (m_{3/2}^2\phi_c/({{\cal V}}M_{\rm pl}))\delta \phi_1 \delta \phi_c$,
$\langle \phi_c \rangle \sim (cHM_*^{n-3})^{1/(n-2)}$ and the inflaton coupling to the modulus is neglected
since it will be dependent on the inflation model. 
As a consequence, for $\langle \phi_c \rangle \lesssim cH M_{\rm pl} m_{\tau_1}{\cal V}/m_{3/2}^2 
\sim cH M_{\rm pl} {\cal V}^{3/2}/\langle W_{\rm eff}\rangle$,
such a picture is valid until $H \sim m_0/c$.
Around $H \sim m_0/c$, the AD fields also starts oscillating and
the effective matrix becomes
\begin{align}
\begin{pmatrix}
m_0^2 & - \frac{m_{3/2}^2}{{\cal V}} \frac{{\langle \phi_c \rangle}}{M_{\rm pl}} \\
- \frac{m_{3/2}^2}{{\cal V}} \frac{{\langle \phi_c \rangle}}{M_{\rm pl}}
 & m_{\tau_1}^2 
\end{pmatrix}
.
\label{Mass2}
\end{align}
Then, so far as 
$\phi_{c, {\rm osc}} \lesssim M_{\rm pl} m_0 m_{\tau_1}{\cal V}/m^2_{3/2} \sim M_{\rm pl}$, 
the off-diagonal components is smaller than the diagonal ones.
Furthermore, with respect to the phase of the MSSM flat direction, the mass mixing between it and the modulus 
is always smaller than the curvature of the phase direction due to the fact of
$\left(\frac{\phi_{c}}{M_{\rm pl}}\right) \ll  1$.
Thus, we conclude that around $H \sim m_0/c $ the true vacuum for the AD fields and the modulus becomes stable;
the AD fields starts oscillating and the baryon asymmetry is generated.

The final baryon asymmetry is written as Eq. (\ref{YB}), assuming  that the lightest modulus dominates 
over the energy density in the Universe. Since the
dilution factor $\Delta^{-1}$ is already computed in the previous subsection, let us consider the typical
magnitude of $c$.
In the geometric regime there will be heavy KK modes and a massive one in the Landau level
on the visible fluxed-brane wrapping on the local 4-cycle $\tau_2$.
They will be interacting with light modes, and the contact term 
$K= |I_c|^2 |\phi_c|^2/\tilde{M}^2$ will be generated after integrating them out.
Thus the mass scale $\tilde M$ is expected as
\begin{align}
\tilde M \sim M_{KK}' \sim \frac{M_{\rm pl}}{\tau_2^{1/4} {\cal V}^{1/2}}
~~~\to~~~
c \sim {\cal V}^{1/2}
.
\end{align}
Note that the condition of Eq. (\ref{condAterm}) is not satisfied in this case.
Thus the resultant baryon asymmetry is given by
\begin{align}
\frac{n_B}{s}(t_0) &\sim 
10^{-10} \delta_{\rm eff} \left(\frac{c}{10^2}\right) 
\left(\frac{T_{\rm dec}^{\phi_1}}{10^3 {\rm GeV} } \right)\left(\frac{m_0}{10^7{\rm GeV}}\right)^{-1} 
\left(\frac{\phi_{c,{\rm osc}}}{10^{14}{\rm GeV}}\right)^2   \bigg(\frac{\Delta \phi_1}{M_{\rm pl}}\bigg)^{-1}\\
& \sim \delta_{\rm eff} \frac{\langle W_{\rm eff} \rangle}{\log({\cal V})^{1/4}{\cal V}} 
\times \bigg(\frac{\Delta \phi_1}{M_{\rm pl}}\bigg)^{-1} 
\sim
\delta_{\rm eff} \frac{1}{\log({\cal V})^{1/4}} \frac{m_{3/2}}{M_{\rm pl}}
\times \bigg(\frac{\Delta \phi_1}{M_{\rm pl}}\bigg)^{-1} 
.
\end{align}
for a numerical example given in the previous subsection. Here we have used the result of
$\phi_{c,{\rm osc}}$ for $n=6$ in $W \sim y\phi^n$.
In this case the baryon would be explained 
for $m_{3/2} = {\cal O}(10^9)$ GeV, ${\cal V}={\cal O}(10^5)$.

Considering the unification of the cutoff scale, 
one might study the case of
\begin{align}
\tilde M \sim M_* \sim M_{\rm pl}/{\cal V}^{1/3} ~\to ~c \sim {\cal V}^{1/3},
\end{align}
though
this operator is understood in terms of the winding mode. 
Again, Eq. (\ref{condAterm}) is not satisfied.
For such a case, one finds
\begin{align}
\frac{n_B}{s}(t_0) &\sim 
10^{-10} \delta_{\rm eff} \left(\frac{c}{10}\right) 
\left(\frac{T_{\rm dec}^{\phi_1}}{10^3 {\rm GeV} } \right)
\left(\frac{m_0}{10^6{\rm GeV}}\right)^{-1} \left(\frac{\phi_{c,{\rm osc}}}{10^{14}{\rm GeV}}\right)^2 
\times \bigg(\frac{\Delta \phi_1}{M_{\rm pl}}\bigg)^{-1} 
\\
&\sim 
\delta_{\rm eff} 
\frac{\langle W_{\rm eff} \rangle}{\log({\cal V})^{1/4}{\cal V}^{7/6}} \times \bigg(\frac{\Delta \phi_1}{M_{\rm pl}}\bigg)^{-1} 
\sim \delta_{\rm eff}
\frac{1}{\log({\cal V})^{1/4}{\cal V}^{1/6}}\frac{m_{3/2}}{M_{\rm pl}} 
\times 
\bigg(\frac{\Delta \phi_1}{M_{\rm pl}}\bigg)^{-1}  .
\end{align}
Thus, in this latter case, the baryon asymmetry will be explained for $m_{3/2} \sim 10^{9}$ GeV
and $\Delta \phi_1 \lesssim M_{\rm pl}$.
Here we have adopted the result for $n=6$, again.

\section{LVS with three K\"ahler moduli in the singular regime}
\label{sec:4}

Here let us study the local model with visible branes on singularities~\cite{Blumenhagen:2009gk},
instead of the geometric regime model.
(For model building, see e.g. \cite{Aldazabal:2000sa}.)
The branes on the singularity lead to a chiral theory with anomalous $U(1)$ symmetries.
We consider the following K\"ahler and super-potentials:
\begin{align}
K &= -2\log(\tau_1^{3/2}-\tau_3^{3/2} + \xi/2) + \frac{(\tau_2 + V_{U(1)})^2}{{\tau_1^{3/2}}} + \frac{|\phi|^2}{\tau_1},
\\
W &= W_{\rm flux} + e^{-aT_3} +W_{\rm MSSM}(\phi),\\
 f_{{\rm vis},a}& = \frac{1}{4\pi} (T_2 + h_a S),
\end{align}
where $V_{U(1)}$ denotes the anomalous U(1) multiplet, $|W_{\rm flux}| = {\cal O}(1)$, which is general  in the flux vacua, 
and $\tau_i = T_i+T^{\dag}_i$.
Thus, by replacing $W_{\rm eff}$ with $W_{\rm flux}$, 
all the results including soft masses are the same as those of the geometric regime, except for $T_2$ and $V_{U(1)}$.
$T_2$ is stabilized via the D-term condition, $D_{U(1)} \sim \partial_{T_2} K \propto \tau_2 =0$, and it is
the Goldstone multiplet absorbed into the $V_{U(1)}$. 
Then one finds
\begin{align}
m_{\tau_2} \sim m_{V_{U(1)}} \sim M_{\rm string} 
\sim \frac{M_{\rm pl}}{{\cal V}^{1/2}} , \qquad
F^{T_2} \propto \partial_{T_2}K  \propto \tau_2 =0.
\end{align}
The result is reasonable since $T_2$ describes the volume of the singularity.
Thus, in order to realize the SM gauge couplings,
 $h_a \langle S \rangle \simeq 25$ is required,
where $h_a$ may take slightly different values depending on the gauge group.
Note also that the cutoff scale is estimated in a similar fashion to the case of the geometric regime,
and it is given by $M_* \sim M_{\rm pl}/{\cal V}^{1/3} \sim M_{\rm wind}$.

Hereafter we focus on the following case,
\begin{align}
{\cal V} = {\cal O}(10^7 -10^8) , \qquad \frac{F^S}{S+S^{\dag}} \sim \frac{1}{{\cal V}^2}
\label{Vvalue}
\end{align}
in order to obtain the moduli masses and SUSY-breaking soft masses which 
are similar to the numerical example  in the geometric regime.
Then one finds 
\begin{align}
M_{a} \sim \frac{F^S}{S+S^{\dag}} \sim {\cal V}^{-2}  \sim 10^2 {\rm\,GeV} - 10 {\rm\,TeV}.
\label{Ma_sing}
\end{align}
For scalar masses, depending on the unknown structure of the matter K\"ahler metric,
they would be suppressed, compared to $\frac{1}{{\cal V}^{3/2}}$.
For instance, given that $Z \sim e^{K_{\rm moduli}/3}$, they almost vanish at the messenger scale.
Hence the resuldant masses are expected to lie in the range of
$M_{1/2} \lesssim m_0 \lesssim \frac{1}{{\cal V}^{3/2}}$\footnote{
After completing this paper, this possibility of suppressed scalar masses was pointed out
during the 3rd UTQuest workshop ExDiP 2012 Superstring Cosmophysics held at
Obihiro in Japan. We thank J.Conlon for pointing out this.
The issue on dark radiation in this context will be further studied in separate papers \cite{Cicoli:2012aq, Higaki:2012ar}.
}
.
Such soft masses will lead to a natural explanation of the presence of
dark radiation originating from
ultralight axions produced by the overall modulus decay
as already mentioned in the footnote \ref{footnoteDR}.
At the moment, we will take $m_0 \sim \frac{1}{{\cal V}^{3/2}}$ for a concreteness.

It is known that, if chiral multiplets charged under the visible gauge group have the $U(1)$ charge,
or if the $U(1)$ interaction is not extremely weak,
there will be 1-loop threshold corrections to the soft masses from 
the heavy gauge multiplet \cite{Shin:2011uk}.
This will  lead to rather heavy gauginos and tachyonic scalars. In order to avoid this,
we neglect those threshold corrections,
assuming that the MSSM fields are not charged under the $U(1)$
and that the MSSM singlet fields charged under the $U(1)$ are heavier than the gravitino via instanton $e^{-2\pi (T_2 + hS)}$.

Let us express the moduli fields in terms of the mass eigenstates as follows \cite{Cicoli:2010ha}:
\begin{align}
\begin{pmatrix}
\delta \tau_1 \\
\delta \tau_2 \\
\delta \tau_3 \\
\delta (1/g_s) 
\end{pmatrix}
\sim 
\begin{pmatrix}
{\cal V}^{2/3} \\
0 \\
\frac{m_{3/2}}{m_{\tau_3}} \\
{\cal V}^{-1/2}
\end{pmatrix}
\delta \phi_1 +
\begin{pmatrix}
0 \\
{\cal V}^{1/2} \\
0 \\
0
\end{pmatrix}
\delta \phi_2 +
\begin{pmatrix}
{\cal V}^{1/6} \\
0 \\
{\cal V}^{1/2} \\
{\cal V}^{-1}
\end{pmatrix}
\delta \phi_3 +
\begin{pmatrix}
{\cal V}^{1/6} \\
0 \\
{\cal V}^{-1/2} \\
{\cal O}(1)
\end{pmatrix}
\delta \phi_s ,
\\
\begin{pmatrix}
\delta \sigma_1 \\
\delta \sigma_2 \\
\delta \sigma_3 \\
\delta C_0^{\rm RR} 
\end{pmatrix}
\sim 
\begin{pmatrix}
{\cal V}^{2/3} \\
0 \\
0 \\
0
\end{pmatrix}
\delta a_1 +
\begin{pmatrix}
0 \\
{\cal V}^{1/2} \\
0 \\
0
\end{pmatrix}
\delta a_2 +
\begin{pmatrix}
{\cal V}^{1/6} \\
0 \\
{\cal V}^{1/2} \\
{\cal V}^{-1}
\end{pmatrix}
\delta a_3 +
\begin{pmatrix}
0 \\
0 \\
{\cal V}^{-1/2} \\
{\cal O}(1)
\end{pmatrix}
\delta a_s .
\end{align}
Note that $\delta \tau_2$ and $\delta \sigma_2$ do not mix with the other fields because of $\tau_2 = 0$.
$\delta a_1$ remains massless as before. 

Suppose that the inflation scale is smaller than or comparable to the lightest modulus mass: $H_{\rm inf} \lesssim m_{\tau_1}$.
Then the lightest modulus $\delta \phi_1$ may induce the cosmological moduli problem, and if so, its decay is relevant for cosmology.
The couplings of the lightest modulus to the visible sector are different
from those in the geometric regime due to the fact 
\begin{align}
\partial_{T_{1,3}} F^{T_2} \sim \partial_{T_{1,3}}\partial_{T_2}K \sim \tau_2 = 0.
\end{align}
There is just a smaller mixing between $T_1$ and $T_2$ than that in the geometric regime.
In particular, the decay into gauginos and higgsino is suppressed as we shall see later.

Instead, 
the moduli-dependent Giudice-Masiero term becomes important:
\begin{align}
\int d^4 \theta \frac{H_u H_d}{(T_1+T^{\dag}_1)} 
& = - 
\int d^4 \theta \frac{\delta T_1^{\dag}}{(T_1+T^{\dag}_1)}\frac{H_u H_d}{(T_1+T^{\dag}_1)} \\
& \supset - \frac{1}{M_{\rm pl}}h_c h_c \partial^2 \delta \phi_1 , \\
 (\delta \tau_1  \partial_{\tau_1} + \delta \tau_3  \partial_{\tau_3})
(B\mu ) H_u H_d & \sim \frac{m_0^2}{M_{\rm pl}} \delta \phi_1 h_c h_c,
\end{align}
where $h_c$ is the canonically normalized (light) Higgs field. The interaction from
the $B \mu$ term is most relevant for the decay of $\delta \phi_1$.
Then the total decay width is given by
\begin{align}
\Gamma_{\phi_1} \simeq \Gamma(\phi_1 \to hh) &\sim \frac{1}{4\pi}\frac{m_{0}^3}{M_{\rm pl}^2}\frac{m_0}{m_{\tau_1}} \\
& \simeq \frac{\log({\cal V})^{2}}{4\pi} \frac{m_{\tau_1}^3}{M_{\rm pl}^2}
\sim \sqrt{\log({\cal V})}\frac{M_{\rm pl}}{{\cal V}^{9/2}},
\end{align}
if the decay is kinematically allowed. 

Let us comment on other decay modes.  Noting that the $\mu$-term is given by
\begin{align}
\mu \simeq m_{3/2} + \frac{F^{T_1}}{2\tau_1}
\sim \frac{1}{\log({\cal V}){\cal V}^2},
\label{mu-term}
\end{align}
 the coupling of $\delta \phi_1$ to the higgsino is suppressed by the $\mu$-parameter:
\beq
\frac{\mu}{M_{\rm pl}}  \delta \phi_1\tilde h_c \tilde h_c.
\eeq
On the other hand, the coupling of $\delta \phi_1$ to gauginos is given by
\beq
\frac{M_{1/2}}{M_{\rm pl}}\delta \phi_1\lambda_c \lambda_c,
\eeq
where we have assumed the dependence of $F^S$ on ${\cal V}$ as $\partial_{\tau_1}F^S \sim m_{3/2}/{\cal V}^{2/3}$.
In addition, although one-loop suppressed, the anomaly mediation  contributes to the coupling to  gauginos.
Thus, the branching fractions of the decay into gauginos and higgsinos are approximately given by
$(M_{1/2}/m_{\tau_1})^2/\log({\cal V})^2 \sim  {\cal V}^{-1}/\log({\cal V})$ and 
$(\mu/m_{\tau_1})^2/\log({\cal V})^2 \sim {\cal V}^{-1}/\log({\cal V})^{3}$, respectively.

In this case that $m_0 \sim 1/{\cal V}^{3/2}$, as in the geometric regime model, 
one can read the branching fraction of $\phi_1 \to 2 a_1$ through the axion kinetic term:
\begin{align}
B(\phi_1 \to a_1 a_1) \simeq \bigg(\frac{m_{\tau_1}}{m_0}\bigg)^4 \sim \frac{1}{\log({\cal V})^2} \sim 3 \times 10^{-3} .
\end{align}
Thus the produced axions from the modulus decay do not give effect on the BBN or CMB again in this case.
Note that the decay to gravitinos is  kinematically forbidden.


The decay temperature of the lightest modulus is given by
\begin{align}
T_{\rm dec}^{\phi_1}   \sim 
6 ~{\rm GeV}\times 
\left(
\frac{m_{\tau_1}}{1.5 \times 10^6{\rm GeV}}
\right)^{3/2}.
\end{align}
Thus the dilution factor by modulus decay will be
\begin{align}
\Delta^{-1} \simeq \frac{T_{\rm dec}^{\phi_1}}{T_{\rm dom}^{\phi_1}} \sim 
2 \times 
10^{-6}\times 
\left(
\frac{m_{\tau_1}}{1.5 \times 10^6{\rm GeV}}
\right)^{3/2}
\left(\frac{T_R}{10^{7}{\rm GeV}}\right)
\left(\frac{\Delta \phi_1}{M_{\rm pl}}\right)^{-1}.
\end{align}
Remember that $\Delta \phi_1/M_{\rm pl} \sim H_{\rm inf}^2/m_{\tau_1}^2$ for
$H_{\rm inf} < m_{\tau_1}$. So, for the modulus mass and the reheating temperature shown in the parenthesis, 
 there is no entropy dilution if the Hubble parameter during inflation is a few orders
of magnitude lower than the modulus mass.

Now let us consider the LSP abundance. 
As one can see from  (\ref{Ma_sing}) and (\ref{mu-term}), the higgsino is likely the LSP.
For ${\cal V} \simeq 10^7$, the LSP mass  is given by $m_\chi \simeq \mu \sim 1$\,TeV, while
the decay temperature is about $200$\,GeV. Thus, soon after the modulus decay, the higgsino LSP 
will be in equilibrium with the ambient plasma.  The right dark matter abundance can be explained
by the thermal relic of the higgsino LSP with mass of about $1$\,TeV~\cite{Hisano:2006nn, ArkaniHamed:2004fb}.
For a smaller value of ${\cal V}$, the LSP abundance exceeds the dark matter density. On the other hand,
for a slightly larger value of ${\cal V}$, it is possible to account for the observed dark matter density by
the non-thermal LSP production by the modulus decay. 

It is possible to make the higgsino heavier by introducing
the dilaton $S$ dependence of the K\"ahler potential as
\begin{align}
\int d^4\theta \frac{|H_u + H_d^{\dag}|^2}{(T_1+T_1^{\dag})} z(S+\bar{S})
\equiv
\int d^4\theta Z {|H_u + H_d^{\dag}|^2}. 
\end{align}
The $\mu$-term is then given by
\begin{align}
\mu \simeq m_{3/2} + {F^I}\partial_I \log (Z) \sim \frac{F^S}{S+S^{\dag}} \sim 
M_a \sim \frac{1}{{\cal V}^2} ={\cal O}(1) ~{\rm TeV}
\end{align}
for ${\cal V}=4.3\times 10^7$.
Alternatively, it is possible to make gauginos lighter by considering
 the case of $F^S \sim (\log({\cal V}){\cal V}^2)^{-1}$ as in the geometric regime.
In these cases,  if the LSP is a certain mixture of the bino and higgsino of mass
${\cal O}(100)$\,GeV, its thermal relic can account for the dark matter~\cite{ArkaniHamed:2006mb}. 
The latest XENON100 result has placed a stringent constraint
on such well-tempered bino-higgsino LSP scenario~\cite{Farina:2011bh}.

Finally let us consider the origin of the effective operator $K= |I_c|^2 |\phi_c|^2/\tilde{M}^2$
and then discuss the result of the AD baryogenesis.
On the visible brane in the singular quiver locus, 
there will be neither KK modes nor Landau level.
However, there are stringy vibration modes.\footnote{Furthermore the anomalous $U(1)$ gauge multiplet would becomes relevant
if both the inflaton and AD fields are charged under the anomalous $U(1)$, 
such an operator is directly generated in this effective theory
after integrating out the heavy multiplet as already mentioned in the section \ref{Kefforigin}. 
However, threshold correction to soft masses also would become relevant.}
Hence one  expects
\begin{align}
\tilde M \sim M_{\rm string} \sim \frac{M_{\rm pl}}{{\cal V}^{1/2}} ~~~\to ~~~
c \sim {\cal V}^{1/2}.
\end{align}
This is similar but slightly different situation with respect to the LVS model in the geometric regime.
Note that Eq. (\ref{condAterm}) is not satisfied in this case.
The final baryon asymmetry is estimated as
\begin{align}
\frac{n_B}{s}(t_0) &\sim 
10^{-10} \delta_{\rm eff} \left(\frac{c}{10^3}\right) 
\left(\frac{T_{\rm dec}^{\phi_1}}{1 {\rm GeV} } \right)
\left(\frac{m_0}{10^7{\rm GeV}}\right)^{-1} 
\left(\frac{\phi_{c,{\rm osc}}}{10^{14}{\rm GeV}}\right)^2
\bigg(\frac{\Delta \phi_1}{M_{\rm pl}}\bigg)^{-1}
\\
& \sim \delta_{\rm eff} \frac{\log({\cal V})^{1/4}}{{\cal V}^{3/2}} \times \bigg(\frac{\Delta \phi_1}{M_{\rm pl}}\bigg)^{-1} 
\sim \delta_{\rm eff} \log({\cal V})^{1/4} \frac{m_{0}}{M_{\rm pl}}
\times \bigg(\frac{\Delta \phi_1}{M_{\rm pl}}\bigg)^{-1} .
\end{align}
For ${\cal V}={\cal O}(10^7 - 10^8)$, $|W_{\rm flux}| ={\cal O}(1)$ and $\Delta \phi_1 \lesssim M_{\rm pl}$,
we obtain the correct baryon asymmetry. Here we have adopted the result for $n=6$ superpotential.

Considering the unification of the cutoff scale again, it is plausible to find the case of
$\tilde M \sim M_* \sim M_{\rm pl}/{\cal V}^{1/3} ~\to ~c \sim {\cal V}^{1/3}$, since 
the winding mode results in generating such a operator.
Again, Eq. (\ref{condAterm}) is not satisfied.
Then, one obtains
\begin{align}
\frac{n_B}{s}(t_0) &\sim
10^{-10} \delta_{\rm eff} \left(\frac{c}{10^2}\right) 
\left(\frac{T_{\rm dec}^{\phi_1}}{1 {\rm GeV} } \right)
\left(\frac{m_0}{10^7{\rm GeV}}\right)^{-1} 
\left(\frac{\phi_{c,{\rm osc}}}{10^{14}{\rm GeV}}\right)^2 \bigg(\frac{\Delta \phi_1}{M_{\rm pl}}\bigg)^{-1} \\
& \sim \delta_{\rm eff} 
\frac{\log({\cal V})^{1/4}}{{\cal V}^{5/3}} \times \bigg(\frac{\Delta \phi_1}{M_{\rm pl}}\bigg)^{-1} 
\sim  \delta_{\rm eff} \frac{\log({\cal V})^{1/4}}{{\cal V}^{1/6}} 
\frac{m_0}{M_{\rm pl}} \times \bigg(\frac{\Delta \phi_1}{M_{\rm pl}}\bigg)^{-1} .
\end{align}
Thus for $m_0 = {\cal O}(10^7)$ GeV, 
the right amount of the baryon asymmetry may be obtained naturally for $\Delta \phi_1 \lesssim M_{\rm pl}$.
Here we have used the result for $n=6$ superpotential, again.

\section{Discussion and Conclusions}
\label{sec:5}
So far we have simply assumed an  inflation model satisfying the constraint (\ref{hinf_const}).
One of such low-scale inflation models is the so called new inflation. For instance, 
we consider the two-field new  inflation model on the visible brane.
The superpotential is given by
\beq
W\;=\; X \left(\mu^2 - \frac{\psi^n}{M^{n-3}} \right),
\eeq
where $\psi$ is the inflaton and $X$ has a non-zero F-term during inflation. 
The Hubble parameter during inflation is of order $\mu^2$. The cut-off scale $M$
is considered to be the winding scale. We assume that the K\"ahler potential is such that
the inflaton mass during inflation is smaller than the Hubble parameter by at least one
order of magnitude. The smallness of $\mu$ can be explained if it arises from the
gaugino condensation or instantons.  Interestingly, it may be possible to suppress the
gravitino production from the inflaton decay~\cite{Kawasaki:2006gs,Asaka:2006bv,Endo:2007ih} in the LVS. This provides another motivation
for the LVS, when the moduli problem is avoided by considering the low-scale
inflation. 

We have also clarified the cosmological moduli problem in a concrete realization of the moduli stabilization.
In particular, we have pointed out that, even if the inflation scale is generically required to be smaller than the modulus mass
in order to avoid decompactification and run-away,  the modulus starts coherent oscillations after inflation with a suppressed amplitude
of order $H_{\rm inf}^2/m_\tau^2$.  This is the case if the inflaton at the potential minimum is heavier than the moduli.
The induced moduli abundance is not negligible unless the Hubble parameter during inflation is many orders of magnitude
smaller than the modulus mass. Furthermore, since the Hubble parameter is bounded below for the AD mechanism to work,
the moduli abundance can be sizable.
Thus, the cosmological moduli problem needs to be considered seriously even if  a
low-scale inflation is assumed.

The recent discovery of the SM-like Higgs boson suggests the high-scale SUSY breaking
at about $10$\,TeV up to the PeV scale. Such high-scale SUSY has the cosmological advantage
of ameliorating the cosmological moduli problem. However, there is  the notorious moduli-induced
gravitino problem: the moduli fields generically decay into gravitinos at a sizable branching fraction. 
Unless the gravitino mass is sufficiently heavy, the gravitino decay produces
too many LSPs, whose abundance easily exceeds the observed dark matter density.
In order to avoid this problem, we have considered a concrete realization of the moduli stabilization in the LVS,
in which the modulus decay into gravitinos is kinematically forbidden. We have shown that the cosmological
moduli problem is indeed solved without the LSP overproduction. 

Another important issue is the baryogenesis. Although the moduli fields decay before the BBN, the pre-existing 
baryon asymmetry is diluted by the huge entropy produced by the modulus decay. Therefore, it is important
to study an efficient baryogenesis mechanism, and we have focused on the AD baryogenesis, taking
also account of possible Q-ball formation. We have shown that the Q balls decay sufficiently fast, both
because the SUSY breaking scale is relatively high, and because of the mild hierarchy
between the scalar mass and the gaugino mass. We have also studied the enhanced coupling between
the inflaton and the AD field, which is expected in the LVS. Interestingly, such an enhanced coupling has turned out to
increase the resultant baryon asymmetry by many orders of magnitude. Furthermore, the enhanced coupling
makes it easy for the AD field to develop a large field value during inflation, which is non-trivial especially
if the inflation scale is bounded above. 

The discovery of the Higgs boson with mass of about $125$\,GeV, therefore, is shedding light not only on the
origin of mass, but also on the beginning of the hot radiation dominated Universe as well as the origin of matter.

\section*{Acknowledgments}
TH would like to thank to H. Abe and T. Kobayashi for fruitful discussions 
on the massive modes in extra dimensions.
This work was supported by the Grant-in-Aid for Scientific Research
on Innovative Areas (No.24111702, No.21111006 and No.23104008) [FT],
Scientific Research (A) (No.22244030 and No.21244033 [FT]), and JSPS
Grant-in-Aid for Young Scientists (B) (No.24740135) [FT].  This work
was also supported by World Premier International Center Initiative
(WPI Program), MEXT, Japan.


\appendix

\section{Example of $c$}

\subsubsection{Examples: Heavy chiral fields}

Consider the K\"ahler potential below with a volume modulus $T = {\cal V}^{2/3}$ 
and the superpotential Eq. (\ref{HSuper}):
\begin{align}
K_{\rm moduli} &= -2 M_{\rm pl}^2\log({\cal V}) = -3 M_{\rm pl}^2\log (T+T^{\dag}), \\
Z_H &= Z_I = Z_\phi = \frac{1}{{\cal V}^{2/3}} =   \frac{1}{(T+T^{\dag})}.
\end{align}
In this case, one finds 
\begin{align}
\tilde{M} = \frac{M_{\rm pl}}{{\cal V}^{1/3}} \sim M_{\rm wind}.
\end{align}
This $\tilde M$ coincides with the physical heavy mass $M_H^{\rm phys} = e^{K/2}M_{\rm pl}/Z$ 
and also the cutoff scale $M_*$ in the physical superpotential
\begin{align}
W = y^{\rm phys}\frac{\phi_c^n}{M_*^{n-3}} = \frac{e^{K/2}y}{\sqrt{Z^3}} \frac{\phi^n }{\sqrt{Z^{n-3}}M_{\rm pl}^{n-3}}.
\end{align}
Note that this is understood as a kind of local models because $y^{\rm phys}$ does not depend on ${\cal V}$.

\subsubsection{Examples: Heavy vector field}

Consider the K\"ahler potential below with two moduli $T = {\cal V}^{2/3}$ 
and $T_v$ Eq. (\ref{HKahler}) and (\ref{HModuli}):
\begin{align}
K_{\rm moduli} &= -2 M_{\rm pl}^2\log({\cal V}) + M_{\rm pl}^2\frac{(T_v + T_v + V_H)^2}{{\cal V}} ,\\
f &= \frac{T_v}{4\pi} + {\rm dilaton}.
\end{align}
Thus it is easy to read the gauge boson mass in the vanishing Fayet-Iliopoulos term $\propto$ $T_v$ limit: 
\begin{align}
\tilde{M} \sim M_H \sim M_{\rm string}  \sim \frac{M_{\rm pl}}{{\cal V}^{1/2}} . 
\end{align}
Here we assumed that the gauge coupling, which will be given by the dilaton, is of $O(1)$. 
If one considers the relevant K\"ahler potential is given by
\begin{align}
K_{\rm moduli} & = - 3 M_{\rm pl}^2\log(T+T^{\dag} + V_H), \\
f &= \frac{T}{4\pi} ,
\end{align}
the D-term condition $D \sim 1/T - q_\psi |\psi_c|^2 \sim 0$ becomes important. 
Then one finds $M_H = g\sqrt{|K'|}  \sim M_{\rm pl}/{\cal V}^{2/3} \sim M_{KK}$ 
and 
\begin{align}
\tilde{M} \sim \sqrt{|K'|} \sim \frac{M_{\rm pl}}{{\cal V}^{1/3}} \sim M_{\rm wind}.
\end{align}

On the other hand,
when the relevant K\"ahler potential is given by
\begin{align}
K_{\rm moduli} & = - 3 M_{\rm pl}^2\log(T+T^{\dag} + V_H) + K_{\rm other ~moduli} (\Phi + \Phi^{\dag} + V_H), \\
f &= \frac{1}{4\pi} \left(T + \sum_i \Phi_i \right),
\end{align}
it will be possible to study the vanishing FI-term, i.e. $\partial_T K +\partial_{\Phi^i}K =0$.
Thus the gauge boson mass will be $M_H \sim g \sqrt{K''} \sim M_{\rm pl}/{\cal V}$ \cite{Cicoli:2011yh}
while 
\begin{align}
\tilde{M} \sim \sqrt{K''} \sim \frac{M_{\rm pl}}{{\cal V}^{2/3}} \sim M_{KK} .
\end{align}
Typical values of $c$ will be exhibited below.

\begin{table}[ht]
\begin{center}
\begin{tabular}{|c|c|c|c|c|} \hline 
$c^2$  & ${\cal V}^{2/3}$ & ${\cal V} $ & ${\cal V}^{4/3} $  \\ \hline \hline
 & $10$      & $30$     &  $10^2$ \\
 & $10^{2}$  & ${10^3}$ &  ${10^4}$  \\
 & $10^{3}$  & ${10^5}$ &  ${10^6}$  \\
 & $10^{9}$ & ${10^{14}}$ & ${10^{18}}$  \\ \hline
\end{tabular}
\caption{Varieties of $c$.
$T \equiv {\cal V}^{2/3}$ is corresponding to the overall volume modulus in supergravity.
Note that gauge coupling at the tree level is given by $g^{-2} \sim T/4\pi $ on the D7-brane in the bulk.}
\end{center}
\end{table}



\end{document}